\RequirePackage{lineno} 
\documentclass[11pt]{elsart}
\usepackage[a4paper]{geometry}
\usepackage{amsfonts,amsmath,amssymb}
\usepackage{natbib, graphicx, amsmath}
\usepackage{url}

\usepackage{cite}
\usepackage{graphicx}
\usepackage{nicefrac}
\usepackage{txfonts}

\newcommand{\addpwd}[1]{#1}

\newcommand{\png}{}       

\newcommand{\PoneDG}{\ensuremath{P_{1\textrm{DG}}}}
\newcommand{\Ptwo}{\ensuremath{P_2}}
\newcommand{\chp}{\PoneDG \Ptwo}
\newcommand{\PDGP}[2]{\ensuremath{P_{#1\textrm{DG}}P_#2}}

\newcommand{\ppt}[1]{\frac{\partial #1}{\partial t}}
\newcommand{\ddt}[1]{\frac{\textrm{d} #1}{\textrm{d} t}}
\newcommand{\ppx}[2][x]{\frac{\partial #2}{\partial #1}}

\newcommand{\eqbreak}{\textrm{\ \ \ }}
\newcommand{\inp}[3]{\left< #1, #2\right>_{#3}}
\newcommand{\Ltwo}{\ensuremath{L_2}}
\newcommand{\norm}[2]{\left| \left| #1 \right| \right|_{#2}}

\renewcommand{\vec}[1]{\mbox{\boldmath$#1$}}

\begin{document}


  \begin{frontmatter}
    \centering \title{Geostrophic balance preserving interpolation in mesh adaptive shallow-water ocean modelling}
    \vspace{-1.0cm}

    \author[ad1]{J. R.\ Maddison\corauthref{cor1}},
    \corauth[cor1]{Corresponding author}
    \ead{maddison@atm.ox.ac.uk}
    \author[ad2]{C. J.\ Cotter},
    \author[ad3,ad4]{P. E.\ Farrell}
    
    \address[ad1]{Atmospheric, Oceanic and Planetary Physics, Department of Physics,\\
      University of Oxford, Oxford, OX1 3PU, UK}
    \address[ad2]{Department of Aeronautics, Imperial College London, London, SW7 2AZ}
    \address[ad3]{Applied Modelling and Computation Group,\\
      Department of Earth Science and Engineering,\\
      Royal School of Mines,\\
      Imperial College London, London, SW7 2AZ, UK}
    \address[ad4]{Institute of Shock Physics,\\
      Royal School of Mines,\\
      Imperial College London, London, SW7 2AZ, UK}

    \journal{Ocean Modelling}
    \begin{abstract}
      The accurate representation of geostrophic balance is an essential
      requirement for numerical modelling of geophysical flows. Significant
      effort is often put into the selection of accurate or optimal
      balance representation by the discretisation of the fundamental
      equations. The issue of accurate balance representation
      is particularly challenging when applying dynamic mesh adaptivity,
      where there is potential for additional imbalance injection when
      interpolating to new, optimised meshes.

      In the context of shallow-water modelling, we present a new
      method for preservation of geostrophic balance when applying
      dynamic mesh adaptivity. This approach is based upon
      interpolation of the Helmholtz decomposition of the Coriolis
      acceleration. We apply this in combination with a discretisation
      for which states in geostrophic balance are exactly steady
      solutions of the linearised equations on an $f$-plane;
      this method guarantees that a balanced and steady flow on a donor
      mesh remains balanced and steady after interpolation onto an
      arbitrary target mesh, to within machine precision. We further
      demonstrate the utility of this interpolant for states close to
      geostrophic balance, and show that it prevents pollution of the
      resulting solutions by imbalanced perturbations introduced by
      the interpolation.
    \end{abstract}

  \end{frontmatter}

\setcounter{section}{0}
\setcounter{equation}{0}
  
\section{Introduction}

It has been recognised in previous work on shallow-water modelling that the
accurate representation of physical balance by the discrete system is of
crucial importance. In \citet{leroux1998} a number of shallow-water finite
element pairs are compared, and it is concluded that the majority of those tested
are unable to represent physical balance accurately while simultaneously remaining
free of spurious numerical modes. In \citet{cotter2009} a
finite element pair is presented using a piecewise
linear discontinuous element for velocity and a piecewise quadratic $C^0$
continuous element for layer thickness: the \chp\ finite element pair.
This is shown to be free of spurious pressure modes
\citep{cotter2009, cotter2009a}
and, in \citet{cotter2009a} in the context of shallow-water modelling, is
shown to have the property that geostrophically balanced states
with a constant stream function on the boundary
are exactly steady solutions of the discrete linearised shallow-water
equations on an $f$-plane. The \chp\ element pair is compared against a number of other low order
discontinuous methods in \citet{comblen2009}, and found to be the most accurate
choice amongst those tested. This discretisation is extended in \citet{cotter2009, cotter2009a}
to form a family of related \PDGP{\left( n - 1 \right)}{n} finite element
pairs, all of which are free of pressure modes and exhibit this optimal balance
property.


This paper is concerned with the application of dynamic mesh adaptivity to
shallow-water ocean modelling. Dynamic mesh adaptivity allows the resolution of the computational mesh to be
varied locally as a simulation develops, in order to resolve dynamically
important regions of the flow and thereby increase the accuracy
per degree of freedom of a model. While this has the potential for enabling
numerical simulations of otherwise inaccessible systems, it presents an
additional problem: as well as the possibility of imbalance injection in the
numerical discretisation of the underlying equations, there is also a potential
for imbalance injection by the mesh optimisation procedure itself.

Recently, dynamic mesh adaptive ocean modelling has been proposed by
\citet{pain2005, piggott2008}. This approach utilises
unstructured meshes in all three dimensions with dynamic mesh adaptivity applied
using element-wise topological operations and nodal perturbations
to optimise the mesh according to a metric derived from the interpolation error
of simulation fields \citep{pain2001, piggott2006, munday2010}. A feature of this approach
is that, in each mesh adapt, the new optimised target mesh has, in general, no
relationship to the original donor mesh, other than that each is some covering
simplex partitioning of the same original domain. This generality
presents a particularly challenging interpolation problem.

The Galerkin projection of fields between arbitrary two and three dimensional meshes
is described in \citet{farrell2009, farrell2009a}. This projection is
(by definition) optimal in the least squares sense, and has the advantage that
the integral of the projected field is exactly conserved. However, there is
no guarantee that the projection injects no additional imbalance. In \citet{stcyr2008} statically refined mesh and uniform mesh
simulations are compared in an adaptive mesh refinement (AMR) shallow-water model
of a geostrophically  balanced flow. It is noted that there is an increase in the
error in the statically refined case, attributed to interpolation errors in the
AMR ghost cells which ``introduce slight disturbance'' in this region,
with the resulting errors growing faster in regions of stronger
solution gradients. It is suggested that higher order schemes be used to mitigate
this.

In this paper we seek to address the problem of imbalance injection by interpolation
between arbitrary unstructured meshes. In the context of shallow-water
modelling, we formulate an interpolant that, for the linearised system on an $f$-plane, 
guarantees that flows that are initially steady and in geostrophic balance
remain steady and in balance after interpolation onto an arbitrary target mesh.

\section{Formulation}

In section \ref{sect:discrete_sw} the finite element discretisation of the
linearised shallow-water equations using the \chp\ finite element pair is
outlined. A set of properties for geostrophic balance preserving
interpolants is presented in section \ref{sect:geostrophic_interpolants} for
which, for the linearised system on an $f$-plane, an initially steady and
geostrophically balanced state remains steady and balanced after
interpolation onto an arbitrary target mesh. An interpolant satisfying these
properties is given in section \ref{sect:implementation}

\subsection{Discretised shallow-water equations}\label{sect:discrete_sw}

The linearised shallow-water equations with free slip boundary conditions are:

\begin{subequations}
  \begin{equation}\label{eqn:sw1}
    \ppt{\vec{u}} + f \vec{\hat{z}} \times \vec{u} + g \nabla \eta = 0,
  \end{equation}
  \begin{equation}\label{eqn:sw2}
    \ppt{\eta} + H \nabla \cdot \vec{u} = 0,
  \end{equation}
  \begin{equation}
    \vec{u} \cdot \vec{\hat{n}} = 0 \textrm{ on } \partial \Omega,
  \end{equation}
\end{subequations}

where $\vec{u}$ is the (horizontal) velocity, $H$ is the mean layer thickness,
$\eta$ is the deviation of the layer thickness from $H$, $f$ is the Coriolis
parameter, $g$ is the gravitational acceleration and $\vec{\hat{n}}$ is a
unit normal on the boundary $\partial \Omega$. From this
one may define two non-dimensional parameters: the Rossby number
$Ro = U / fD$ and the Froude number $Fr = \sqrt{U / gH}$,
where $U$ and $D$ are characteristic flow speeds and spatial scales
respectively.

Multiplying equations \eqref{eqn:sw1} and \eqref{eqn:sw2} by test functions
$\vec{w}$ and $\phi$ respectively, integrating over the domain
$\Omega$, integrating by parts and applying the free slip boundary
condition yields the weak form:

\begin{subequations}
  \begin{equation}\label{eqn:sw1_weak}
    \int_\Omega \vec{w} \ppt{\vec{u}} + f  \vec{w} \cdot \left(\vec{\hat{z}} \times \vec{u} \right) + g \vec{w} \cdot \nabla \eta = 0 \eqbreak \forall \vec{w},
  \end{equation}
  \begin{equation}\label{eqn:sw2_weak}
    \int_\Omega \phi \ppt{\eta} - H \int_\Omega \nabla \phi \cdot \vec{u} = 0 \eqbreak \forall \phi.
  \end{equation}
\end{subequations}

Choosing some simplex covering partition of $\Omega$ (the mesh), restricting
$\vec{w}$ and $\vec{u}$ to be piecewise linear discontinuous and restricting
$\phi$ and $\eta$ to be piecewise quadratic $C^0$ continuous completes
the \chp\ spatial discretisation:

\begin{subequations}
  \begin{equation}
    \int_\Omega \vec{w}^\delta \ddt{\vec{u}^\delta} + f  \vec{w}^\delta \cdot \left(\vec{\hat{z}} \times \vec{u}^\delta \right) + g \vec{w}^\delta \cdot \nabla \eta^\delta = 0 \eqbreak \forall \vec{w}^\delta,
  \end{equation}
  \begin{equation}
    \int_\Omega \phi^\delta \ddt{\eta}^\delta - H \int_\Omega \nabla \phi^\delta \cdot \vec{u}^\delta = 0 \eqbreak \forall \phi^\delta,
  \end{equation}
\end{subequations}

where $\zeta^\delta$ denotes the finite element approximation for $\zeta$.
Introducing basis function expansions of $\vec{w}^\delta$, $\vec{u}^\delta$,
$\phi^\delta$ and $\eta^\delta$, this can be re-expressed as:

\begin{subequations}
  \begin{equation}\label{eqn:sw1_discrete}
    \ddt{} M_1 \vec{\tilde{u}} + f L \vec{\tilde{u}} + g C \tilde{\eta} = 0,
  \end{equation}
  \begin{equation}\label{eqn:sw2_discrete}
    \ddt{} M_2 \tilde{\eta} - H C^T \vec{\tilde{u}} = 0,
  \end{equation}
\end{subequations}

where $\vec{\tilde{u}}$ and $\tilde{\eta}$ are the nodal values for velocity
and layer thickness respectively and:

\begin{align}
  M_1 = \textrm{diag} \left( M_1', M_1'\right), & & M_2 = \textrm{diag} \left( M_2', M_2'\right), \nonumber \\
  L = \left( \begin{array}{cc}0 & -1 \\ 1 & 0\end{array}\right) M_1', & & C_T = \left( C^x, C^y \right).
\end{align}

$M_1'$ and $M_2'$ are the velocity space and layer thickness space mass matrices
respectively, and $C^T$ is the discrete divergence matrix:

\begin{align}
  \left( M_1' \right)_{ij} = \int_\Omega \psi_i \psi_j, & & \left( M_2' \right)_{ij} = \int_\Omega \xi_i \xi_j, \nonumber
\end{align}
\begin{equation}\label{eqn:matrices}
  \left( C^q \right)_{ij} = \int_\Omega \ppx[q]{\xi_i} \psi_j \eqbreak q \in \{x, y\}, 
\end{equation}

where the $\psi_i$ and $\xi_i$ are the \PoneDG\ and \Ptwo\ elemental basis functions
respectively. Choosing some time discretisation allows equations \eqref{eqn:sw1_discrete}
and \eqref{eqn:sw2_discrete} to be integrated on a computer.

\subsection{Geostrophic balance preserving interpolants}\label{sect:geostrophic_interpolants}

Consider interpolation between a donor mesh $A$ and a target mesh $B$.
Let $\left( \ldots \right)^A$ and $\left( \ldots \right)^B$ denote ``on donor''
and ``on target'' respectively - for example $(C^T)^A$ and $(C^T)^B$ are the
divergence matrices, as per \eqref{eqn:matrices}, assembled
on the donor and target meshes respectively. Consider an interpolation procedure
as follows:

\begin{enumerate}
  \item Perform a Helmholtz decomposition of the Coriolis acceleration
        $\vec{F_*}^A = f \vec{\hat{z}} \times \vec{\tilde{u}}^A$ on
        the donor mesh:
        \begin{equation}\label{eqn:discrete_helmholtz_a}
          M_1^A \vec{F}^A = M_1^A \vec{F_*}^A + C^A \Phi^A,
        \end{equation}
        for some scalar potential $\Phi^A$ and discrete divergence
        free $\vec{F}^A$.
  \item Interpolate $\vec{F}^A$, $\Phi^A$ and $\tilde{\eta}^A$ from the
        donor to the target to form $\vec{F}^B$, $\Phi^B$ and
        $\tilde{\eta}^B$, using interpolants with the following
        properties:
    \begin{subequations}
      \begin{equation}\label{eqn:bal_bound}
        \left( M_1^A \right)^{-1} C^A \Phi^A \times \vec{\hat{n}} = 0 \textrm{ on } \partial \Omega \Rightarrow \left( M_1^B \right)^{-1} C^B \Phi^B \times \vec{\hat{n}} = 0 \textrm{ on } \partial \Omega,
      \end{equation}
      \begin{equation}\label{eqn:bal_kernel}
        \vec{F}^A = 0 \Rightarrow \vec{F}^B = 0,
      \end{equation}
      \begin{equation}\label{eqn:bal_pressure}
        \Phi^A = g \tilde{\eta}^A \Rightarrow \Phi^B = g \tilde{\eta}^B.
      \end{equation}
    \end{subequations}
  \item Recompose $\vec{F_*}^B$ from $\Phi^B$ and $\vec{F}^B$:
    \begin{equation}\label{eqn:discrete_helmholtz_b}
      M_1^B \vec{F_*}^B = M_1^B \vec{F}^B - C^B \Phi^B.
    \end{equation}
\end{enumerate}

The Helmholtz decomposition splits the Coriolis acceleration into a curl-free
scalar potential gradient component and a divergence free
residual \citep{weyl1940, ladyzhenskaya1969}. Only the scalar potential gradient component can be
cancelled from equation \eqref{eqn:sw1_discrete} by a layer thickness gradient.
For incompressible Navier-Stokes, this scalar potential gradient component must be exactly
cancelled by the pressure gradient, with the diagnostic pressure field acting as
a Lagrange multiplier via which the incompressibility constraint is applied.

Property \eqref{eqn:bal_kernel} states the somewhat trivial requirement that zero
is preserved by the interpolant for $\vec{F}$. Property \eqref{eqn:bal_pressure}
couples the velocity and layer thickness interpolation, and can be
achieved if $\Phi$ and $\tilde{\eta}$ use the same interpolant and that
interpolant is scale invariant. Property \eqref{eqn:bal_bound} asserts that
the interpolant for $\Phi$ preserves zero tangential gradients on the
domain boundary, and is required in order to avoid generation of grid
scale boundary Kelvin waves by the interpolation.

We now proceed to prove that this interpolant, when applied to a \chp\
discretisation of the linearised shallow-water equations on an
$f$-plane, guarantees that a steady and geostrophically balanced state
on the donor mesh results in a state that is steady and balanced on the target mesh. By
definition, for a geostrophically balanced state on the donor:

\begin{equation}\label{eqn:balance_a}
  f L^A \vec{\tilde{u}}^A = -g C^A \tilde{\eta}^A.
\end{equation}

By equation \eqref{eqn:discrete_helmholtz_a} $\vec{F}^A = 0$ and
$\Phi^A = g \tilde{\eta}^A$, and hence by properties \eqref{eqn:bal_kernel} and
\eqref{eqn:bal_pressure} $\vec{F}^B = 0$ and $\Phi^B = g \tilde{\eta}^B$.
Hence, by equation \eqref{eqn:discrete_helmholtz_b}, on the target:

\begin{equation}\label{eqn:balance_b}
  f L^B \vec{\tilde{u}}^B = -g C^B \tilde{\eta}^B.
\end{equation}

Also, since $\vec{F_*}$ is perpendicular to $\vec{\tilde{u}}$:

\begin{align}\label{eqn:interp_bound}
  \vec{\tilde{u}}^A \cdot \vec{\hat{n}} = 0 \textrm{ on } \partial \Omega & \Rightarrow \vec{F_*}^A \times \vec{\hat{n}} = 0 \textrm{ on } \partial \Omega & \nonumber \\
                                                                          & \Rightarrow (M_1^A)^{-1}C^A \Phi^A \times \vec{\hat{n}} = 0 \textrm{ on } \partial \Omega & \textrm{ by \eqref{eqn:balance_a}} \nonumber \\
                                                                          & \Rightarrow (M_1^B)^{-1}C^B \Phi^B \times \vec{\hat{n}} = 0 \textrm{ on } \partial \Omega & \textrm{ by \eqref{eqn:bal_bound}} \nonumber \\
                                                                          & \Rightarrow \vec{F_*}^B \times \vec{\hat{n}} = 0 \textrm{ on } \partial \Omega & \nonumber \\
                                                                          & \Rightarrow \vec{\tilde{u}}^B \cdot \vec{\hat{n}} = 0 \textrm{ on } \partial \Omega. 
\end{align}

From \citep{cotter2009a}, if using the \chp\ element pair on an $f$-plane:

\begin{align}\label{eqn:chp_magic}
  f L \vec{\tilde{u}} + g C \tilde{\eta} = 0 \textrm{ and } \vec{\tilde{u}} \cdot \vec{\hat{n}} = 0 \textrm{ on } \partial \Omega \nonumber \\
    \Leftrightarrow \ppt{\vec{\tilde{u}}} = 0 \textrm{ and } \ppt{\tilde{\eta}} = 0.
\end{align}

Hence by \eqref{eqn:balance_b}, \eqref{eqn:interp_bound} and 
\eqref{eqn:chp_magic} the solution on the target mesh is geostrophically
balanced and exactly steady.

\subsection{Implementation and boundary conditions}\label{sect:implementation}

The Helmholtz decomposition of the Coriolis acceleration on the donor mesh
is equivalent to the pressure projection method commonly used for
incompressible Navier-Stokes solvers \citep{chorin1967, temam1968, gresho1990}.
Multiplying equation \eqref{eqn:discrete_helmholtz_a}  by
$(C^T)^A (M_1^A)^{-1}$ and using $C^T \vec{F}^A = 0$ leads to the
elliptic equation:

\begin{equation}\label{eqn:poisson}
  (C^T)^A (M_1^A)^{-1} C^A \Phi^A = - (C^T)^A \vec{F_*}^A.
\end{equation}

Note that here the consistent mass matrix can be used as the \PoneDG\
mass matrix $M_1^A$ is block diagonal, and hence the Laplacian matrix
$(C^T)^A (M_1^A)^{-1} C^A$ is sparse. From this $\Phi^A$ and $\vec{F}^A$
can be determined. Following interpolation of the Helmholtz decomposition
$\vec{F_*}^B$ can be diagnosed directly from $\Phi^B$
and $\vec{F}^B$ via equation \eqref{eqn:discrete_helmholtz_b}.
Therefore, the key step in forming
a geostrophic balance preserving interpolant is to choose interpolants
for $\Phi$, $\tilde{\eta}$ and $\vec{F}$ such that the properties
\eqref{eqn:bal_bound}, \eqref{eqn:bal_kernel} and
\eqref{eqn:bal_pressure} are satisfied.

One simple approach is to apply a Galerkin projection of $\vec{F}$ from the
donor mesh to the target mesh, as described in \citet{farrell2009}, and to
interpolate $\Phi$ and $\tilde{\eta}$ using collocation: evaluation
of the donor fields at the nodal coordinates of the target mesh. Since Galerkin
projection and collocation are linear, properties \eqref{eqn:bal_kernel}
and \eqref{eqn:bal_pressure} are satisfied. Collocation also
preserves constant boundary values, and hence property \eqref{eqn:bal_bound} is
satisfied. However, collocation (at least for piecewise linear fields)
erodes solutions bounds and has no optimality properties \citep{farrell2009a}.

A more accurate approach is to apply a mesh-to-mesh Galerkin projection of $\Phi$ and
$\tilde{\eta}$. This does not in general satisfy property \eqref{eqn:bal_bound}, although this
issue can be resolved by using a further decomposition of $\Phi$, with an
equivalent decomposition of $\tilde{\eta}$ in order to satisfy property
\eqref{eqn:bal_pressure}. Assuming $\Omega$ is simply connected, $\Phi^A$
can be decomposed into:

\begin{equation}\label{eqn:p_decomp}
  \Phi^A = \Phi_C^A + \Phi_R^A,
\end{equation}

where $\Phi_C^A$ is equal to some constant $c$ on the boundary, and
$\Phi_R^A$ is some residual. $\Phi_C^A$ can be re-expressed:

\begin{equation}
  \Phi_C^A = \Phi_0^A + c \Phi_1.
\end{equation}

Here $\Phi_0$ and $\Phi_1$ satisfy:

\begin{subequations}
  \begin{equation}\label{eqn:p_decomp_solve1}
    N^A_0 \Phi_0^A = N^A \Phi^A,
  \end{equation}
  \begin{equation}\label{eqn:p_decomp_solve2}
    N^A_1 \Phi_1^A = 0,
  \end{equation}
\end{subequations}

where $N^A$ is the discrete Laplacian matrix
$N^A = (C^T)^A (M_1^A)^{-1} C^A$, $N_0^A$ is $N^A$ with a Dirichlet
boundary condition of zero on $\partial \Omega$ and $N_1^A$ is $N^A$ with a
Dirichlet boundary condition of one on $\partial \Omega$.
Minimising $\norm{\Phi_R^A}{L^2}$
subject to \eqref{eqn:p_decomp} then yields a unique value for $c$:

\begin{equation}\label{eqn:p_decomp_bound}
  c = \frac{\inp{\Phi_1^A}{\Phi^A - \Phi_0^A}{L^2}}{\inp{\Phi_1^A}{\Phi_1^A}{L^2}}.
\end{equation}

This choice of $c$ has the property that if $\Phi^A$ is constant on
$\partial \Omega$, then $\Phi_R^A=0$. Applying a Galerkin projection of $\Phi_C^A$ and
$\Phi_R^A$ from the donor mesh to the target mesh, with a Dirichlet boundary
condition of $c$ on $\partial \Omega$ for $\Phi_C^A$, therefore guarantees
that the boundary property \eqref{eqn:bal_bound} is satisfied.

Note that using the mass matrix in place of the Laplacian matrix, $N^A = M_2^A$,
is not suitable here, as this results in non-smooth $\Phi_C^A$ and $\Phi_R^A$,
with strong gradients close to the boundary which can
generate significant noise in the donor-to-target Galerkin projection.

%
%

The full geostrophic balance preserving interpolation procedure is therefore:

\begin{enumerate}
  \item Compute $\vec{F_*}^A$ from $\vec{\tilde{u}}^A$.
  \item Solve equation \eqref{eqn:poisson} for $\Phi^A$ and compute
        $\vec{F}^A$ using equation \eqref{eqn:discrete_helmholtz_a}.
  \item Solve equations \eqref{eqn:p_decomp_solve1} and \eqref{eqn:p_decomp_solve2},
        with $N^A = (C^T)^A (M_1^A)^{-1} C^A$, for $\Phi_0^A$ and $\Phi_1^A$,
        and compute $\Phi_C^A$ and $\Phi_R^A$.
        Perform a similar decomposition for $\tilde{\eta}^A$ to form $\tilde{\eta}_C^A$
        and $\tilde{\eta}_R^A$.
  \item Apply a Galerkin projection from the donor mesh to the target mesh of
        $\Phi_C^A$, $\Phi_R^A$, $\tilde{\eta}_C^A$, $\tilde{\eta}_R^A$
        and $\vec{F}^A$, with Dirichlet boundary conditions for
        $\Phi_C^A$ and $\tilde{\eta}_C^A$ as determined from equation \eqref{eqn:p_decomp_bound},
        to form $\Phi_C^B$, $\Phi_R^B$, $\tilde{\eta}_C^B$, $\tilde{\eta}_R^B$
        and $\vec{F}^B$.
  \item Compute $\Phi^B$ from $\Phi_C^B$ and $\Phi_R^B$ using equation \eqref{eqn:p_decomp}.
        Similarly compute $\tilde{\eta}^B$ from $\tilde{\eta}_C^B$ and $\tilde{\eta}_R^B$.
  \item Compute $\vec{F_*}^B$ using equation \eqref{eqn:discrete_helmholtz_b}.
  \item Compute $\vec{\tilde{u}}^B$ from $\vec{F_*}^B$.      
\end{enumerate}

\section{Numerical Examples}

In this section several numerical examples of geostrophic balance
preservation using the interpolation procedure presented above are
given. In section \ref{sect:balance_preservation} it is demonstrated that
the geostrophic balance preserving interpolant ensures that a steady
and balanced state remains steady and balanced after interpolation.
In section \ref{sect:nearly_balanced} a state close to geostrophic
balance is considered, and it is shown that the geostrophic balance
preserving interpolant avoids imbalance injection. The interpolant
is applied to a Kelvin wave in section \ref{sect:kelvin_wave},
and the accuracy of the interpolant in the \Ltwo\ norm is quantified
in section \ref{sect:accuracy}.

\subsection{Preservation of balance}\label{sect:balance_preservation}

The \chp\ linearised shallow-water equations \eqref{eqn:sw1_discrete} and
\eqref{eqn:sw2_discrete} on an $f$-plane were discretised in time using Crank-Nicolson
finite differencing \citep{crank1947}, and the linear systems solved with
preconditioned conjugate gradients using the PETSc 
library \citep{balay1997, petsc-manual, petsc-web-page}.
Further details of the discretisation are given in \citet{cotter2009a}.

\begin{figure}
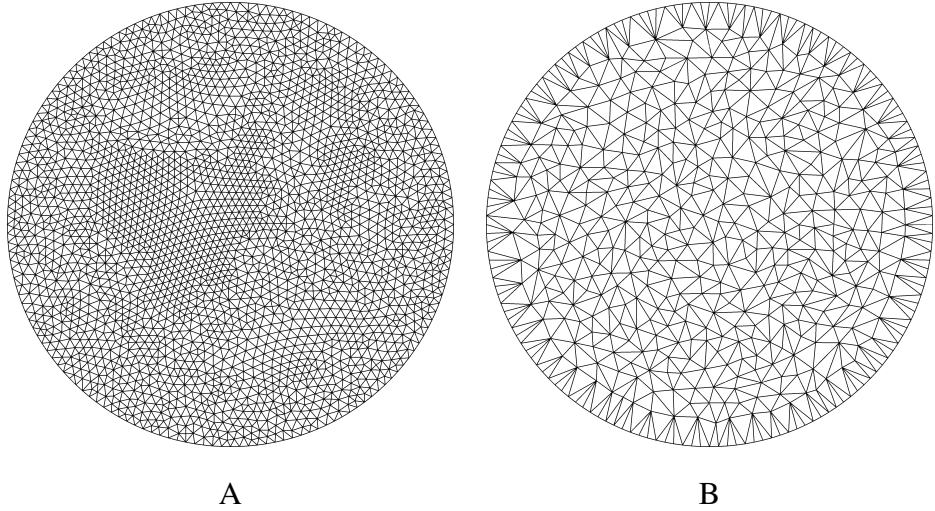

  \begin{centering}
    \begin{tabular}{c c}
      \includegraphics[width=0.4\textwidth]{\addpwd{BalancedMeshA\png}}
    & \includegraphics[width=0.4\textwidth]{\addpwd{BalancedMeshB\png}} \\
    A & B \\
    \end{tabular}
    \caption{Pseudo-isotropic meshes used to test for imbalance injection by mesh-to-mesh
    	     interpolation. Mesh B has one half the resolution of mesh A.}\label{fig:balance_meshes}
  \end{centering}
\end{figure}

\begin{figure}
  \begin{centering}
    \includegraphics[width=0.5\textwidth]{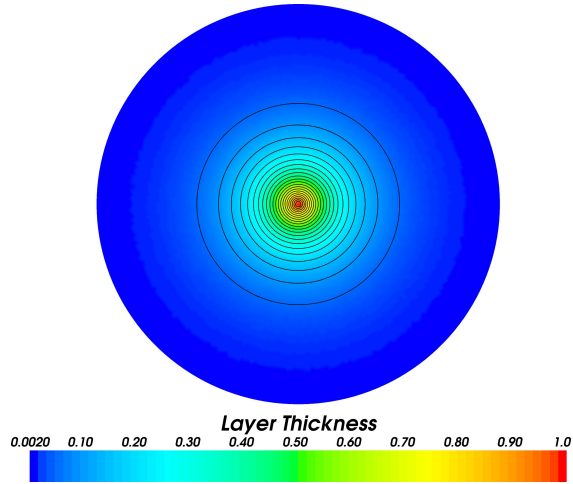}
    \caption{Gaussian profile layer thickness used as an initial condition for
             the geostrophic balance preservation test.}\label{fig:balance_initial}
  \end{centering}
\end{figure}

In order to test for imbalance injection by mesh-to-mesh interpolation,
two pseudo-isotropic circular meshes A and B were generated using Gmsh
\citep{geuzaine2009} and the ani2d mesh optimisation library
\citep{vasilevskii1999, agouzal1999}, with mesh A of one half the
resolution of mesh B, as shown in figure \ref{fig:balance_meshes}.
Meshes A and B have $2447$ and $557$ nodes respectively.
Following the balance preservation test of \citet{leroux1998, cotter2009},
the system was initialised on mesh A with a Gaussian profile for layer
thickness, shown in figure \ref{fig:balance_initial}, and with a velocity field
initialised so as to be in discrete geostrophic balance with that profile
as per equation \eqref{eqn:balance_a}. The solution was then interpolated
backwards and forwards between meshes A and B at ten timestep intervals.
Since geostrophically balanced states with a constant stream function
on the boundary are known to be exactly steady when
using the \chp\ element pair \citep{cotter2009a}, any transience
observed in the simulation is purely due to imbalance injection by the
interpolation procedure.

\begin{figure}
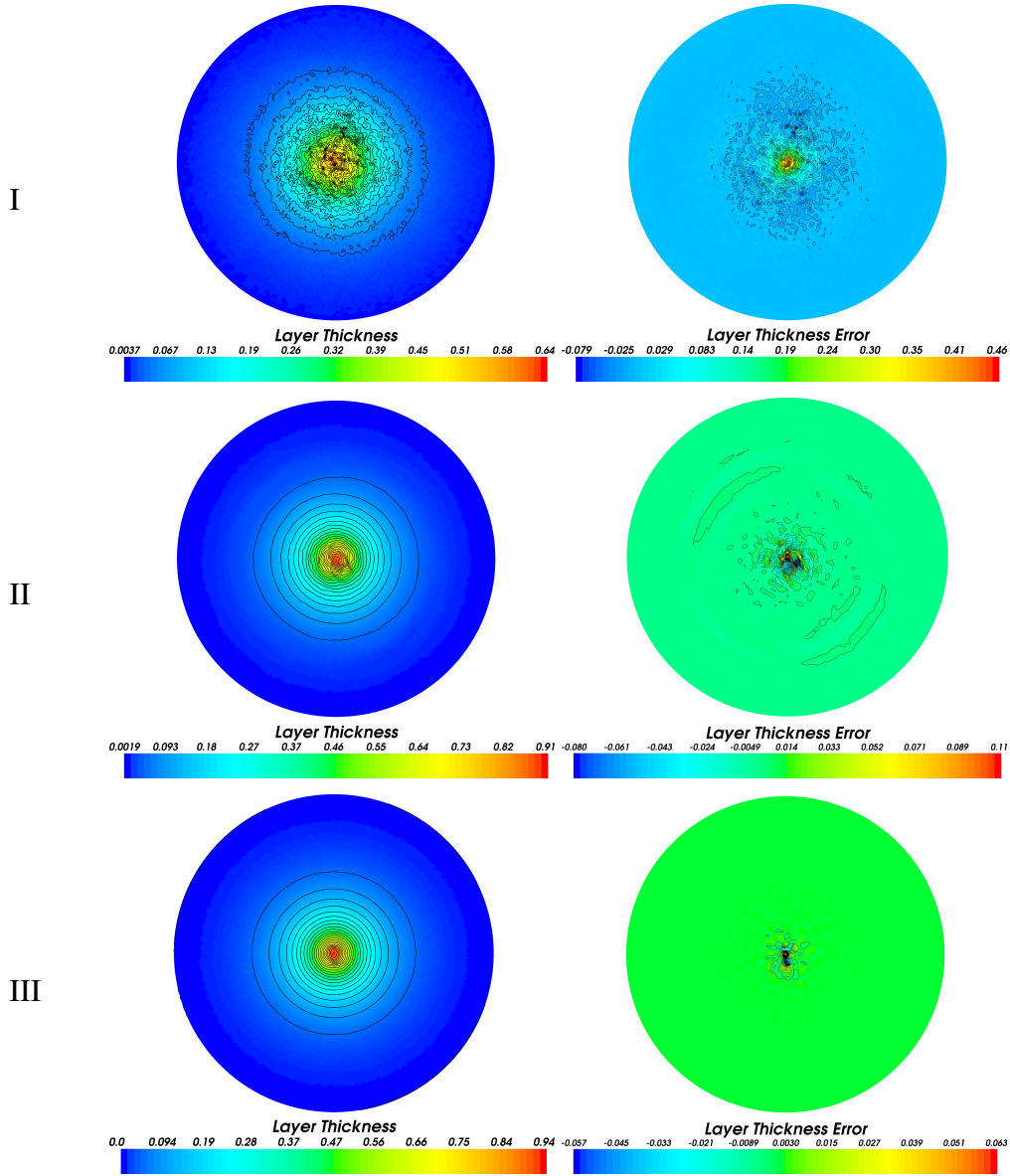

  \begin{centering}
    \begin{tabular}{p{0.7cm} c}
     I & 
      \begin{tabular}{c c}
        \includegraphics[width=0.4\textwidth]{\addpwd{BalancedFinalLayerThicknessGrandy\png}}
        \includegraphics[width=0.4\textwidth]{\addpwd{BalancedFinalLayerThicknessErrorGrandy\png}}
      \end{tabular} \\
     II & 
      \begin{tabular}{c c}
        \includegraphics[width=0.4\textwidth]{\addpwd{BalancedFinalLayerThicknessGp\png}}
        \includegraphics[width=0.4\textwidth]{\addpwd{BalancedFinalLayerThicknessErrorGp\png}}
      \end{tabular} \\
     III & 
      \begin{tabular}{c c}
        \includegraphics[width=0.4\textwidth]{\addpwd{BalancedFinalLayerThicknessGs\png}}
        \includegraphics[width=0.4\textwidth]{\addpwd{BalancedFinalLayerThicknessErrorGs\png}}
      \end{tabular}
    \end{tabular}
    \caption{The final solution of the geostrophic balance preservation
             test after 20 repeated interpolations backwards and forwards
             between the pseudo-isotropic meshes A and B in figure
             \ref{fig:balance_meshes}.
             Left: Final layer thickness.
             Right: Change in layer thickness from the initial condition
             in figure \ref{fig:balance_initial}.
             I: Grandy interpolation.
             II: Galerkin projection.
             III: Helmholtz decomposed geostrophic balance preserving
                interpolation.
             }\label{fig:balance_final}
  \end{centering}
\end{figure}

\begin{figure}
  \begin{centering}
    \includegraphics[width=0.8\textwidth]{\addpwd{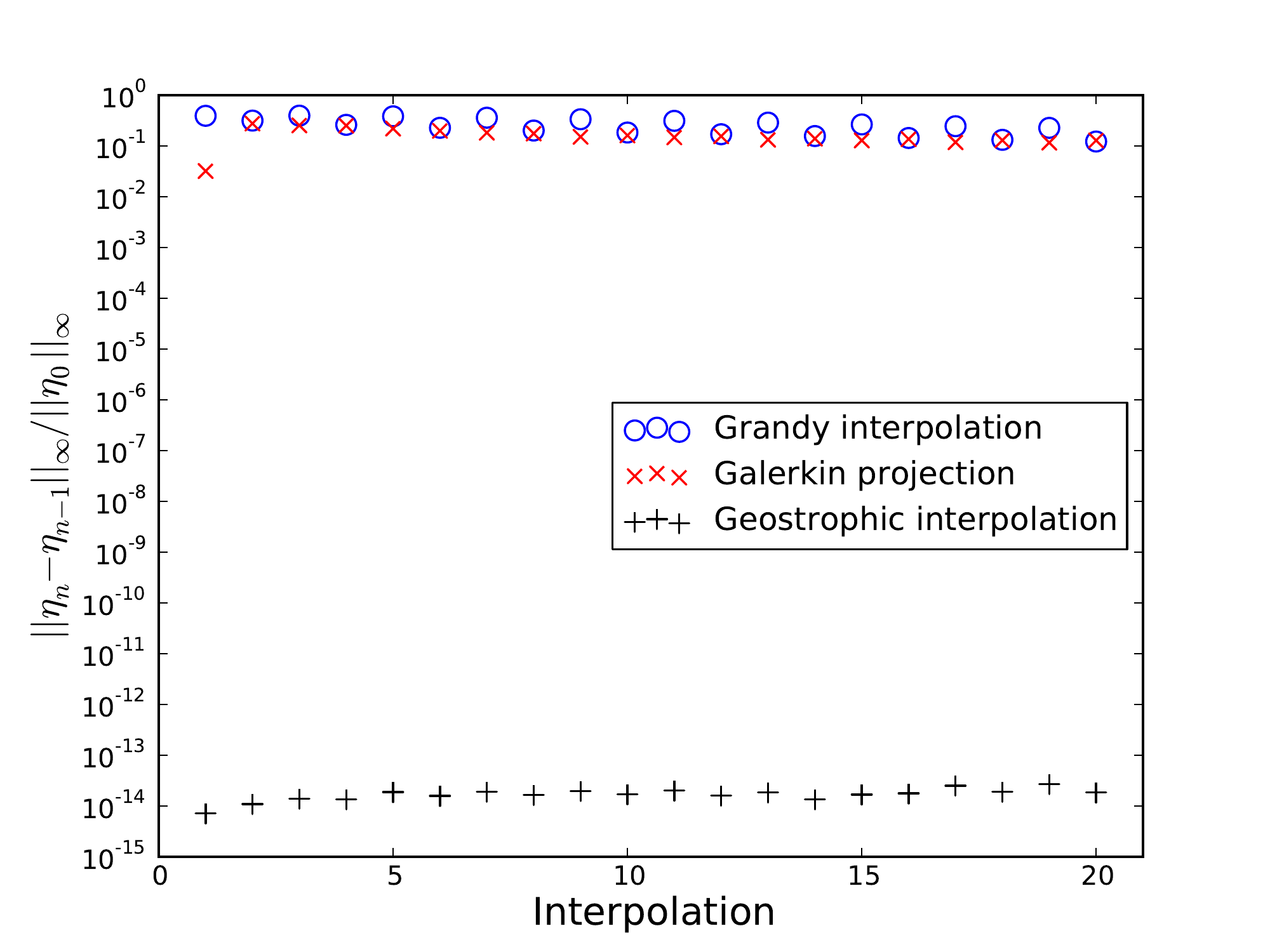}}
    \caption{The maximum change in layer thickness between interpolations
             backwards and forwards between the pseudo-isotropic meshes A and B
             in figure \ref{fig:balance_meshes}.}\label{fig:balance_lt_change}
  \end{centering}
\end{figure}


The model was integrated for $210$ timesteps of $6 \times 10^{-4} (D / U)$
at Rossby number $0.06$ and Froude number $0.07$ for a
total of $20$ interpolations between meshes A and B. Three interpolants were tested: a first order
accurate interpolant as proposed by \citet{grandy1999} for both velocity
and layer thickness (``Grandy interpolation''), Galerkin projection for
velocity and layer thickness, and the geostrophic balance preserving
interpolant presented in the previous section. Collocation was not tested, as
this is not well defined at element boundaries and hence is unsuitable for use
with the discontinuous velocity field.

The final layer thickness
and change in layer thickness from the initial condition are shown
in figure \ref{fig:balance_final}, and the maximum change in layer thickness
between each interpolation is shown in figure \ref{fig:balance_lt_change}.
Grandy interpolation is observed to inject imbalance everywhere after each
interpolation, resulting in a severe degradation of the
simulation fields after just a single interpolation. Galerkin projection is
observed to inject imbalance towards the centre of the domain, near the
layer thickness maximum. The resulting gravity waves propagate outwards 
polluting the global solution, and accumulate after every interpolation.
The geostrophic balance preserving interpolant is exactly steady, to
within machine precision, after every interpolation, with a change in layer
thickness between interpolations of $\lesssim10^{-13}$. The residual imbalance
between interpolations is attributed to double precision round-off error.

After $20$ interpolations the \Ltwo layer thickness error is $20\%$ (of
initial layer thickness \Ltwo\ norm) for Grandy
interpolation, $2.7\%$ for Galerkin projection and $2.0\%$ for the geostrophic
balance preserving interpolant. While Galerkin projection is optimal in
the \Ltwo\ norm for each interpolation, the imbalance injection and
resulting pollution of the solution by gravity waves leads to a reduced
accuracy in the \Ltwo\ norm of the final model solution with respect
to the geostrophic balance preserving interpolant.

\begin{figure}
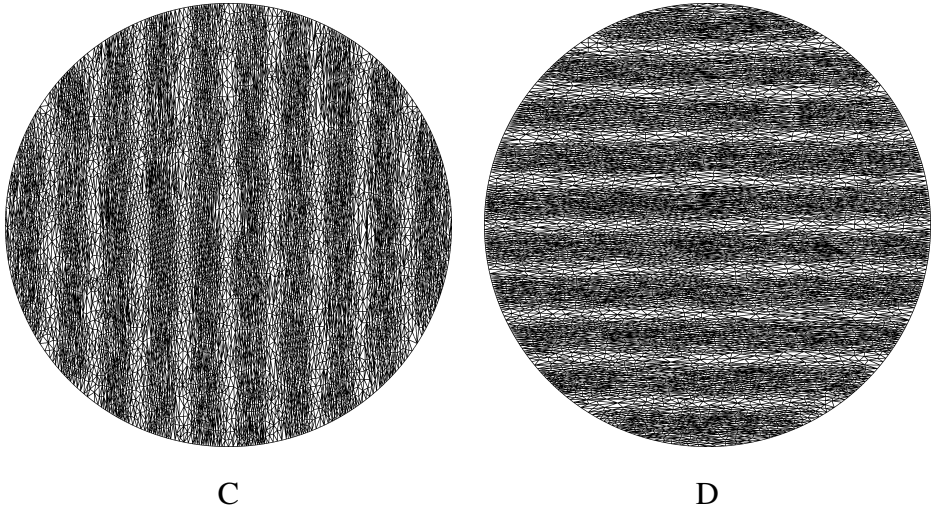

  \begin{centering}
    \begin{tabular}{c c}
      \includegraphics[width=0.4\textwidth]{\addpwd{BalancedAnisotropicMeshA\png}}
    & \includegraphics[width=0.4\textwidth]{\addpwd{BalancedAnisotropicMeshB\png}} \\
    C & D \\
    \end{tabular}
    \caption{Anisotropic meshes used to test for imbalance injection by mesh-to-mesh
    	     interpolation. There is no relationship between meshes C and D,
             other than that they cover the same domain.}\label{fig:balance_meshes_anisotropic}
  \end{centering}
\end{figure}

\begin{figure}
  \begin{centering}
    \includegraphics[width=0.8\textwidth]{\addpwd{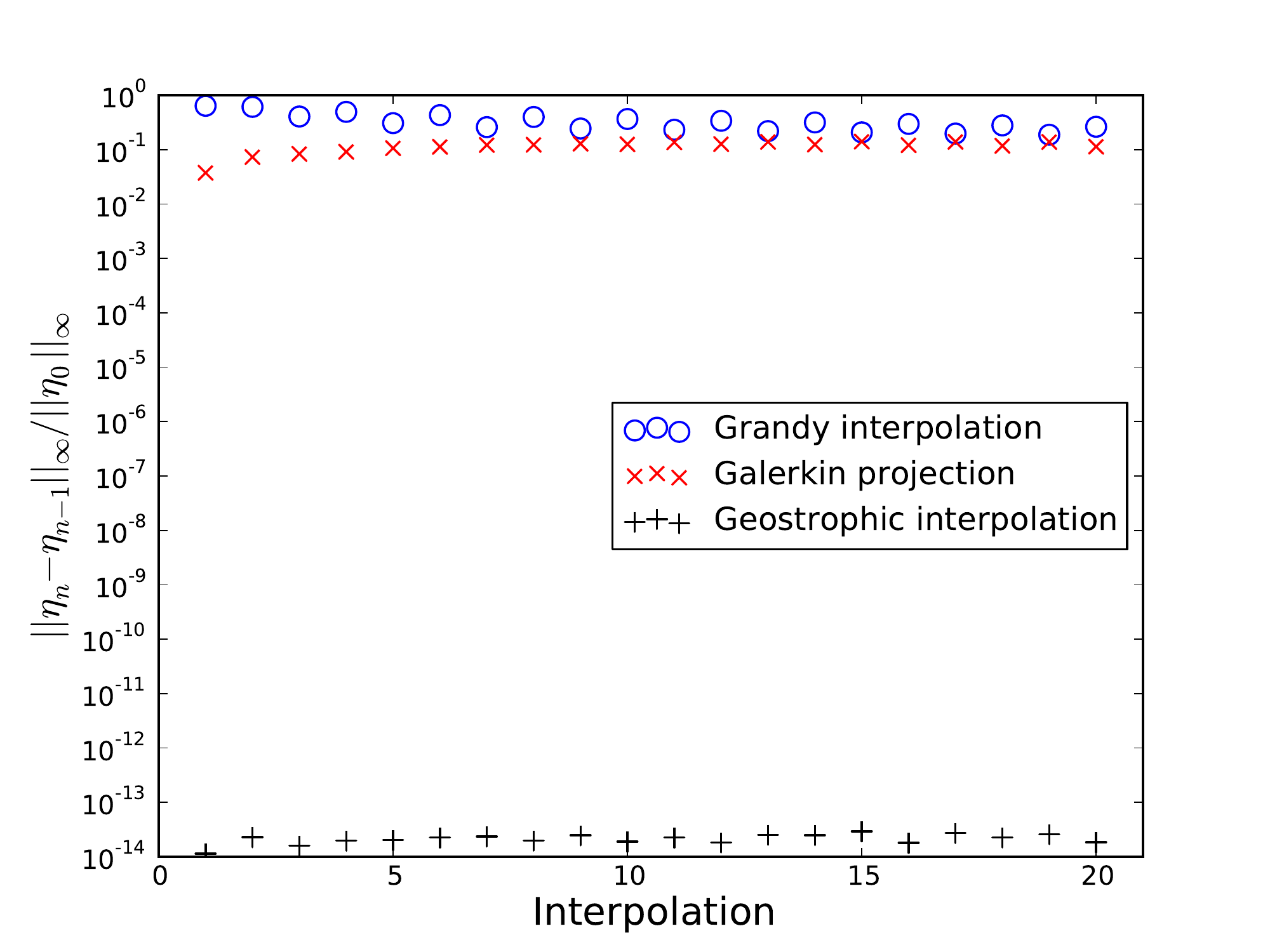}}
    \caption{The maximum change in layer thickness between interpolations
             backwards and forwards between the anisotropic meshes C and D
             in figure \ref{fig:balance_meshes_anisotropic}.}\label{fig:balance_lt_change_anisotropic}
  \end{centering}
\end{figure}

To further demonstrate geostrophic balance preservation the test was repeated
on two \linebreak[4] anisotropic circular meshes C and D generated using Gmsh
\citep{geuzaine2009} and the ani2d mesh optimisation library
\citep{vasilevskii1999, agouzal1999} with elements stretched in
perpendicular directions as shown in figure \ref{fig:balance_meshes_anisotropic}. 
Meshes C and D have $7986$ and $7205$ nodes respectively, and a maximum element
edge length ratio of $\sim30$. The velocity field was initialised to be
in discrete geostrophic balance with this layer thickness as before, with
interpolations backwards and forwards between the two meshes at $10$ timestep
intervals for $20$ interpolations. Simulations were conducted using the
geostrophic balance preserving interpolant, Galerkin projection, and Grandy
interpolation, with the change in layer thickness between interpolations shown
in figure \ref{fig:balance_lt_change_anisotropic}.
When applying the geostrophic balance preserving
interpolation the maximum change between interpolations was $\lesssim 10^{-13}$
as before.

Finally, the geostrophic balance preservation the test was repeated using the
anisotropic meshes C and D in figure \ref{fig:balance_meshes_anisotropic},
with a layer thickness initialised to random values in the interior and a value
of zero on the boundary.
The velocity field was initialised to be in discrete geostrophic balance
with this layer thickness. The model was integrated as before, with
interpolations backwards and forwards between two two meshes at $10$
timestep intervals for $4$ interpolations, using the geostrophic balance
preserving interpolant. The maximum change in layer thickness between
interpolations was $\lesssim 10^{-12}$, and hence the solution
was observed to be steady to within double precision round-off error.

\subsection{Nearly balanced states}\label{sect:nearly_balanced}

\begin{figure}
  \begin{centering}
    \includegraphics[width=0.5\textwidth]{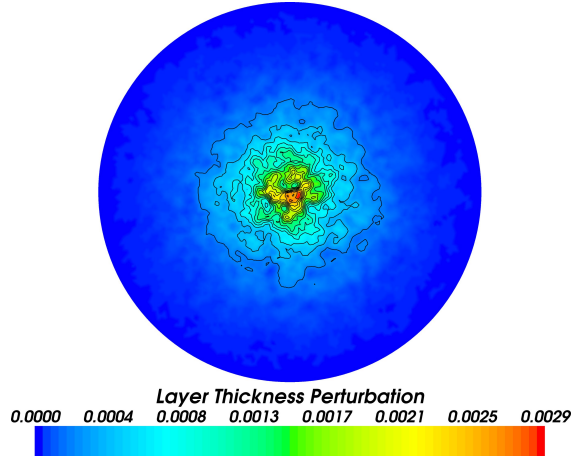}
    \caption{Perturbation applied to the layer thickness in figure \ref{fig:balance_initial}
             to test for imbalance injection by interpolation of a nearly balanced state.}\label{fig:nearly_balanced_initial_perturbation}
  \end{centering}
\end{figure}

The Gaussian layer thickness profile in figure \ref{fig:balance_initial} had
a perturbation applied of the form:

\begin{equation}
  \Delta \tilde{\eta} = \frac{1}{10} X \tilde{\eta},
\end{equation}

where $X$ is some point-wise random value in the range $\{ 0 - 1 \}$. This perturbation was
smoothed using a Helmholtz smoother with a characteristic length scale of
$D / 8$ to produce the layer thickness perturbation shown in figure \ref{fig:nearly_balanced_initial_perturbation}.
The velocity field was initialised to be in discrete geostrophic balance
with the unperturbed layer thickness, thereby generating a nearly balanced
state.

\begin{figure}
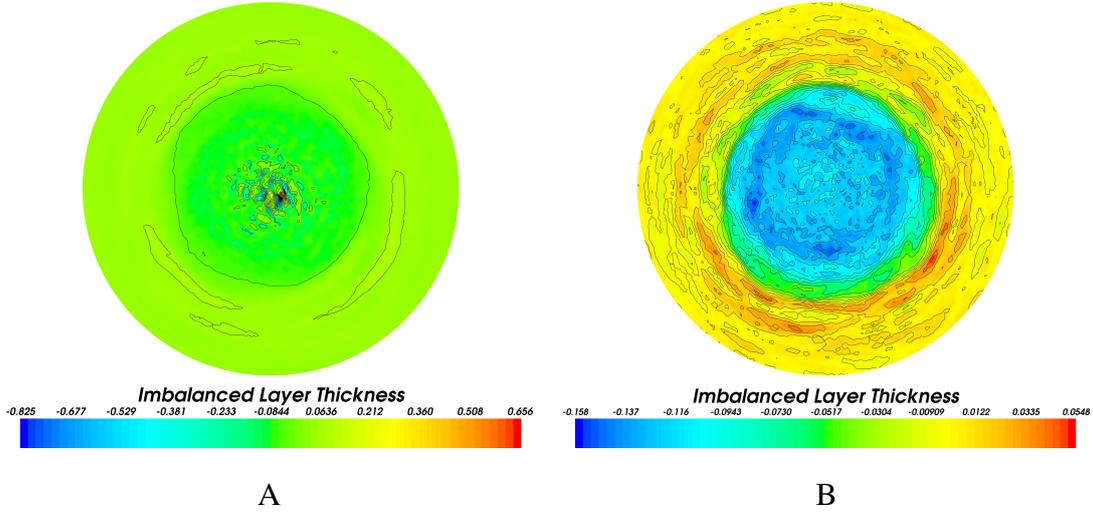

  \begin{centering}
    \begin{tabular}{c c}
      \includegraphics[width=0.47\textwidth]{\addpwd{NearlyBalancedFinalImbalancedLayerThicknessGp\png}} &
      \includegraphics[width=0.47\textwidth]{\addpwd{NearlyBalancedFinalImbalancedLayerThicknessGs\png}} \\
    A & B \\
    \end{tabular}
    \caption{The final imbalanced layer thickness, computed as the difference
             between the simulation layer thickness and the scaled scalar
             potential from the Helmholtz decomposition of the Coriolis
             acceleration.
             A: Galerkin projection.
             B: Helmholtz decomposed geostrophic balance preserving
                interpolation.}\label{fig:nearly_balanced_final_perturbation}
  \end{centering}
\end{figure}

The system was integrated as before, with interpolations backwards and
forwards between the pseudo-isotropic meshes A and B in figure
\ref{fig:balance_meshes} at $10$ timestep intervals. One can define an
``imbalanced layer thickness'':

\begin{equation}
  \tilde{\eta}_{imbal} \coloneqq∗ \tilde{\eta} - \frac{1}{g} \Phi,
\end{equation}

where $\Phi$ is the scalar potential computed from the Helmholtz decomposition of the
Coriolis acceleration. The final imbalanced layer thickness is shown for Galerkin
projection and the geostrophic balance preserving interpolant in figure
\ref{fig:nearly_balanced_final_perturbation}. When using Galerkin projection
imbalance is observed to be injected near the layer thickness maximum. This
additional imbalance dominates over the original layer thickness perturbation
after $20$ interpolations.
When using the geostrophic balance preserving interpolant propagation of the
original layer thickness perturbation is observed, with no significant
pollution introduced by the interpolation.

Defining a ``balanced velocity'' $\vec{\tilde{u}}_{bal}$ where:

\begin{equation}
  f L \vec{\tilde{u}}_{bal} \coloneqq∗ - g C \tilde{\eta},
\end{equation}

and an ``imbalanced velocity''  $\vec{\tilde{u}}_{imbal}$:

\begin{equation}
  \vec{\tilde{u}}_{imbal} \coloneqq∗ \vec{\tilde{u}} - \vec{\tilde{u}}_{bal},
\end{equation}

allows one to compute an imbalanced kinetic energy:

\begin{align}
  T_{imbal} & = \half \norm{\vec{\tilde{u}}_{imbal}}{L^2}^2 \nonumber \\
            & = \half \norm{\vec{\tilde{u}} + \frac{g}{f} L^{-1} C \tilde{\eta}}{L^2}^2.
\end{align}

\begin{figure}
  \begin{centering}
    \includegraphics[width=0.8\textwidth]{\addpwd{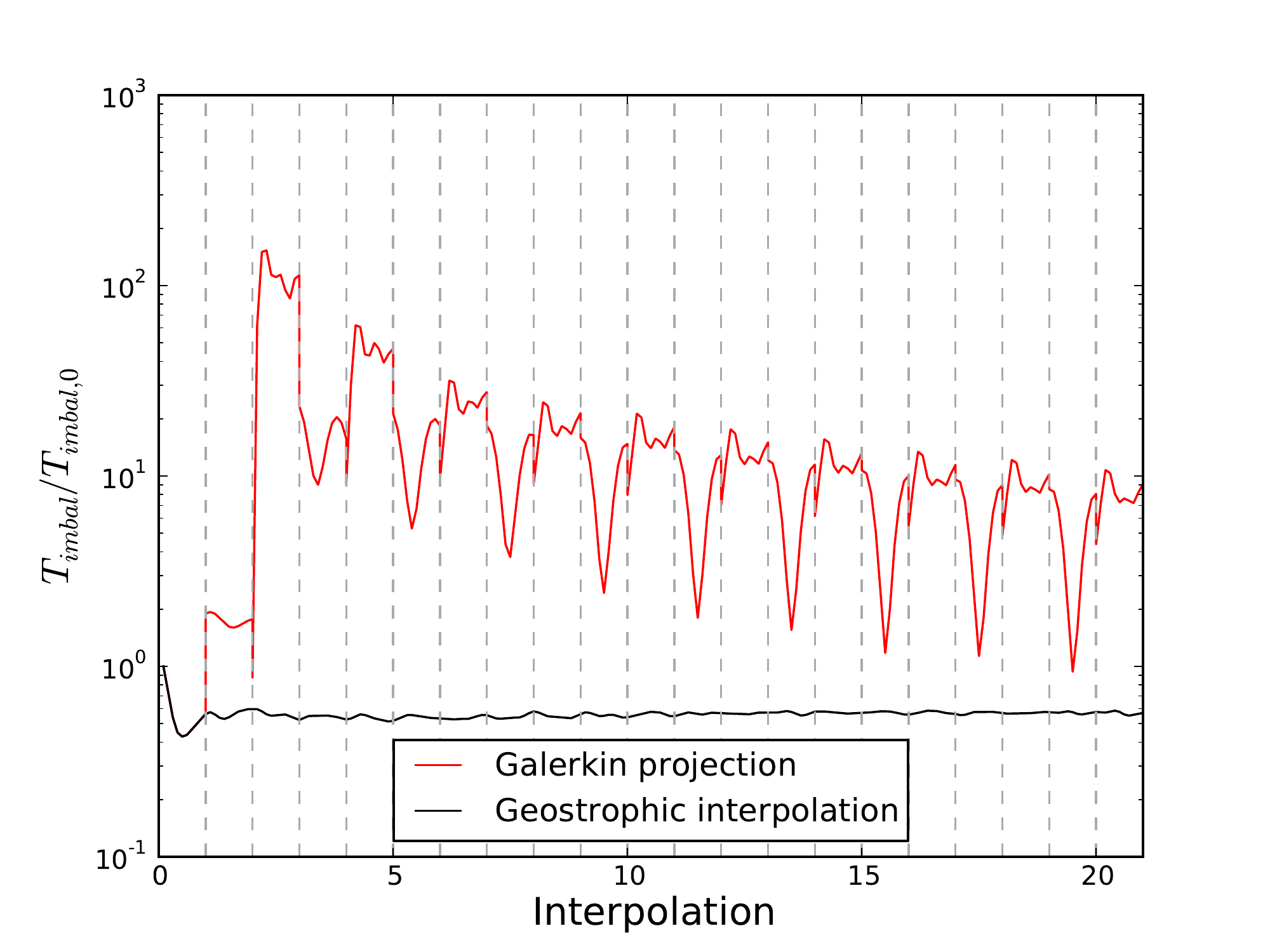}}
    \caption{The imbalanced kinetic energy, normalised by the initial
             imbalanced kinetic energy, for the nearly balanced interpolation
             test. When using Galerkin projection (upper line) large
             increases in the imbalanced kinetic energy are observed after interpolation,
             with these increases dominating over the original imbalanced kinetic energy.
             When using the geostrophic balance preserving interpolant (lower line)
             imbalanced kinetic energy injection is significantly reduced.}\label{fig:nearly_balanced_ke}
  \end{centering}
\end{figure}

The imbalanced kinetic energies when using Galerkin projection and the \linebreak[4]
geostrophic balance preserving interpolant are shown in
figure \ref{fig:nearly_balanced_ke}.
When using Galerkin projection the imbalanced kinetic energy is observed
to increase by up to a factor of $70$ in an interpolation, with the
imbalanced kinetic energy peaking at $150$ times its initial value.
When using the geostrophic balance preserving interpolant the imbalanced kinetic
energy is observed to increase by at most a factor $1.02$ in an
interpolation, and the imbalanced kinetic energy never exceeds
its initial value.

\subsection{Kelvin wave}\label{sect:kelvin_wave}

The interpolant was tested for a Kelvin wave, configured
as in in \citet{ham2005, cotter2009} with an initial condition:

\begin{subequations}
  \begin{equation}
    \eta \left( r, \theta \right) = \exp{\left( \frac{r - r_0}{Ro} \right)} \cos{\theta},
  \end{equation}
  \begin{equation}
    u_\theta \left( r, \theta \right) = \frac{1}{Fr} \exp{\left( \frac{r - r_0}{Ro} \right)} \cos{\theta},
  \end{equation}
  \begin{equation}
    u_r = 0,
  \end{equation}
\end{subequations}

for $Ro = 10$ and $Fr = 1$ in a circular domain of radius $r_0$. The
Kelvin wave is geostrophically balanced in the direction normal to the boundary
and imbalanced in the tangential direction. The model was integrated with
a timestep of $2 \pi \times 10^{-4} (D / U)$ for a total simulated time of
$2 \pi (D / U)$, corresponding to the time taken for a single Kelvin
wave to perform a circuit of the domain in the limit of large $r_0$.
Two meshes of quasi-uniform resolution with $1473$ and $1461$ nodes
respectively were created using gmsh \citep{geuzaine2009} and the ani2d
mesh optimisation library \citep{vasilevskii1999, agouzal1999}, and the solution
interpolated backwards and forwards between these meshes at $10$ timestep intervals.

\begin{figure}
    \begin{centering}
    \includegraphics[width=0.5\textwidth]{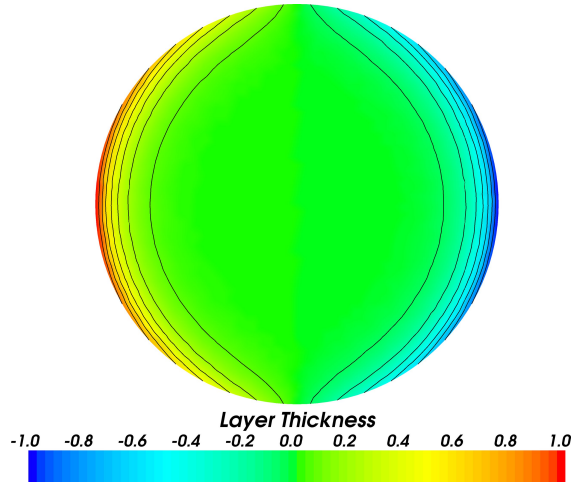}
    \caption{Initial layer thickness used for the Kelvin wave test at
             $Ro = 10$, $Fr = 1$.}\label{fig:kw_initial}
  \end{centering}
\end{figure}

\begin{figure}
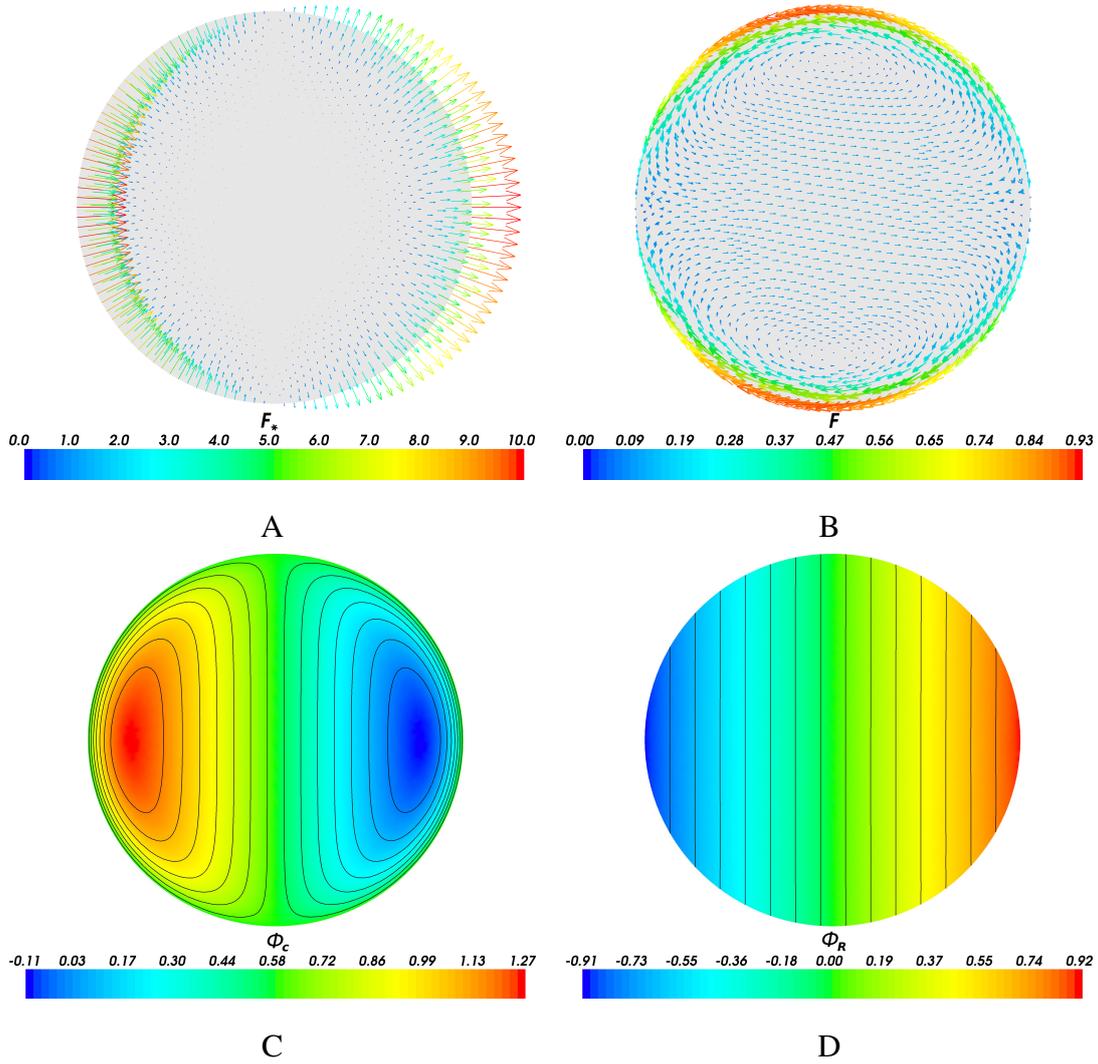

  \begin{centering}
    \begin{tabular}{c c}
      \includegraphics[width=0.47\textwidth]{\addpwd{KwInitialCoriolis\png}} &
      \includegraphics[width=0.47\textwidth]{\addpwd{KwInitialF\png}} \\
    A & B \\
      \includegraphics[width=0.47\textwidth]{\addpwd{KwInitialPhiC\png}} &
      \includegraphics[width=0.47\textwidth]{\addpwd{KwInitialPhiR\png}} \\
    C & D \\
    \end{tabular}
    \caption{Helmholtz decomposition of initial Coriolis acceleration for the
             Kelvin wave test at $Ro = 10$, $Fr = 1$.
             A: Coriolis acceleration, $\vec{F_*}$.
             B: Non-divergent residual, $\vec{F}$.
             C: Scalar potential with a constant boundary value, $\Phi_c$.
             D: Scalar potential residual, $\Phi_R$.}\label{fig:kw_initial_helmholtz}
  \end{centering}
\end{figure}

\begin{figure}
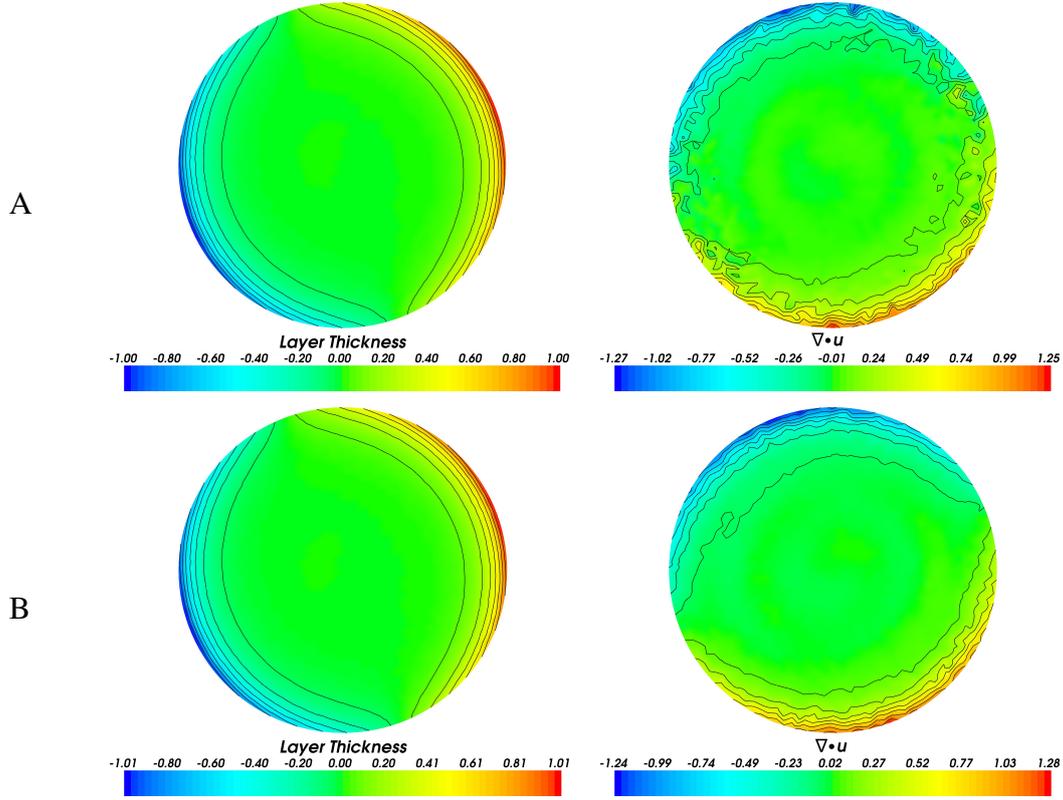

  \begin{centering}
    \begin{tabular}{p{0.7cm} c}
     A & 
      \begin{tabular}{c c}
        \includegraphics[width=0.41\textwidth]{\addpwd{KwFinalLayerThicknessGp\png}} &
        \includegraphics[width=0.41\textwidth]{\addpwd{KwFinalDivUGp\png}}
      \end{tabular} \\
     B & 
      \begin{tabular}{c c}
        \includegraphics[width=0.41\textwidth]{\addpwd{KwFinalLayerThicknessGs\png}} &
        \includegraphics[width=0.41\textwidth]{\addpwd{KwFinalDivUGs\png}}
      \end{tabular}
    \end{tabular}
    \caption{The final solution of the Kelvin wave test
             at $Ro = 10$, $Fr = 1$
             Left: Final layer thickness.
             Right: Final velocity divergence, $M_2^{-1} C^T \vec{\tilde{u}}$.
             A: Galerkin projection.
             B: Helmholtz decomposed geostrophic balance preserving
                interpolation.}\label{fig:kw_final}
  \end{centering}
\end{figure}

The initial layer thickness is shown in figure \ref{fig:kw_initial}, and the Helmholtz decomposition
of the initial Coriolis acceleration in figure \ref{fig:kw_initial_helmholtz}. 
The final solutions when using Galerkin projection and
the geostrophic balance preserving interpolant are shown in figure \ref{fig:kw_final}.
Relatively little difference is observed in the final layer thickness field between
these simulations.
However, when using Galerkin projection, noise is observed in the velocity divergence
field, originating at the boundary.
This noise is significantly reduced when using the geostrophic balance preserving
interpolant.

\begin{figure}
    \begin{tabular}{p{0.7cm} c}
     A & 
      \begin{tabular}{c}
        \includegraphics[width=0.46\textwidth]{\addpwd{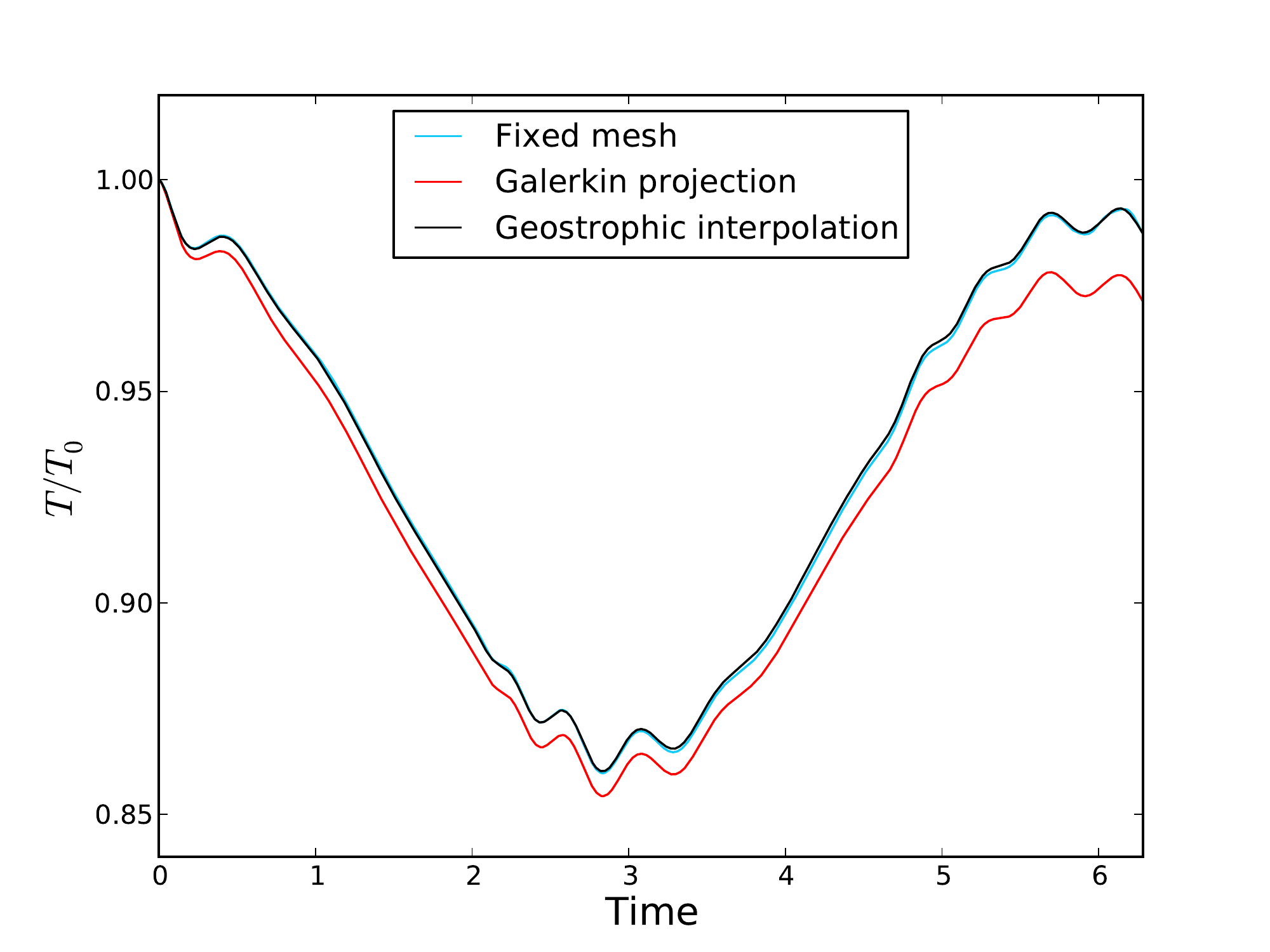}} 
      \end{tabular}
      \\
     B & 
      \begin{tabular}{c}
        \includegraphics[width=0.46\textwidth]{\addpwd{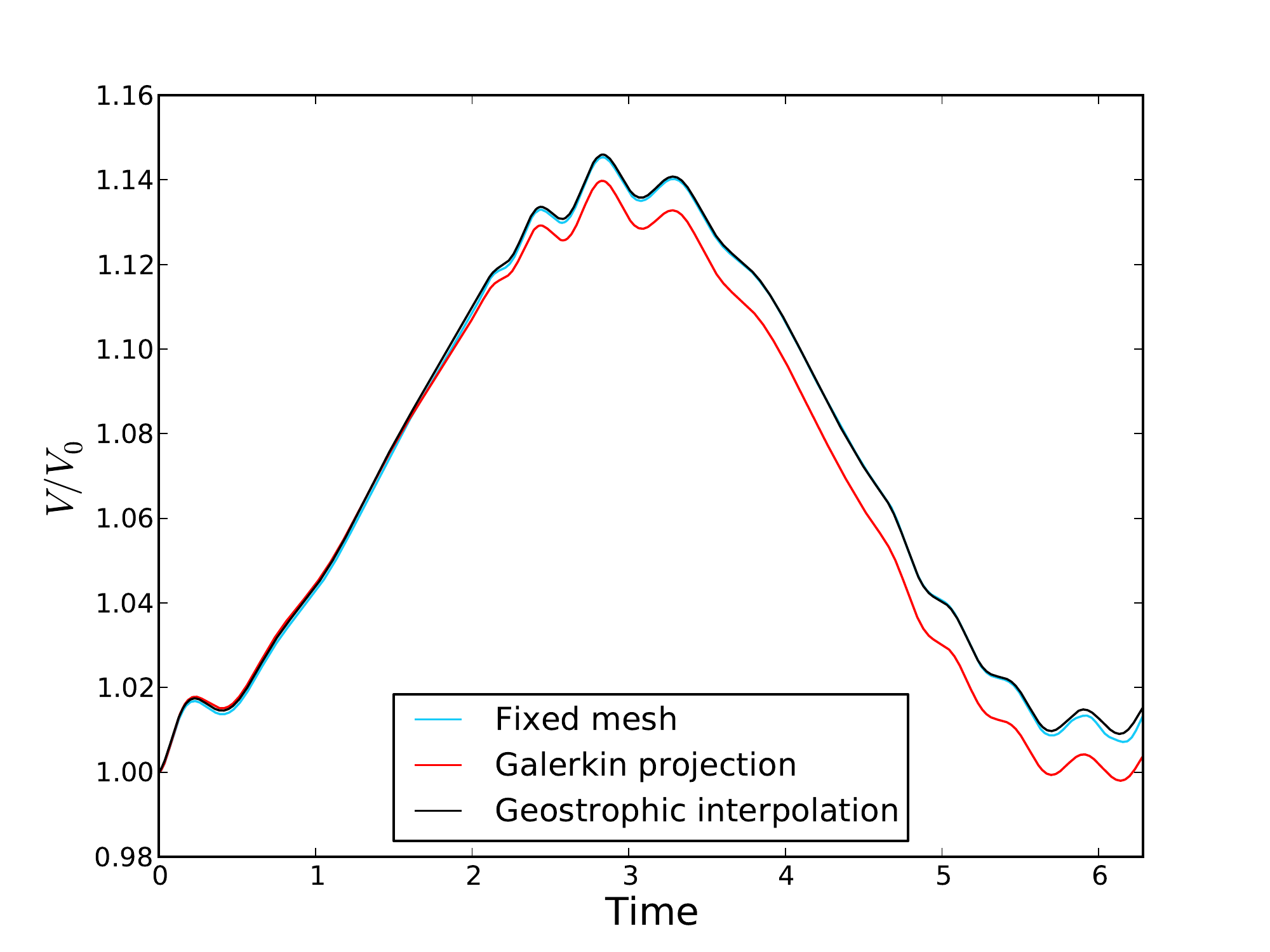}}
      \end{tabular}
      \\
     C &
      \begin{tabular}{c}
        \includegraphics[width=0.46\textwidth]{\addpwd{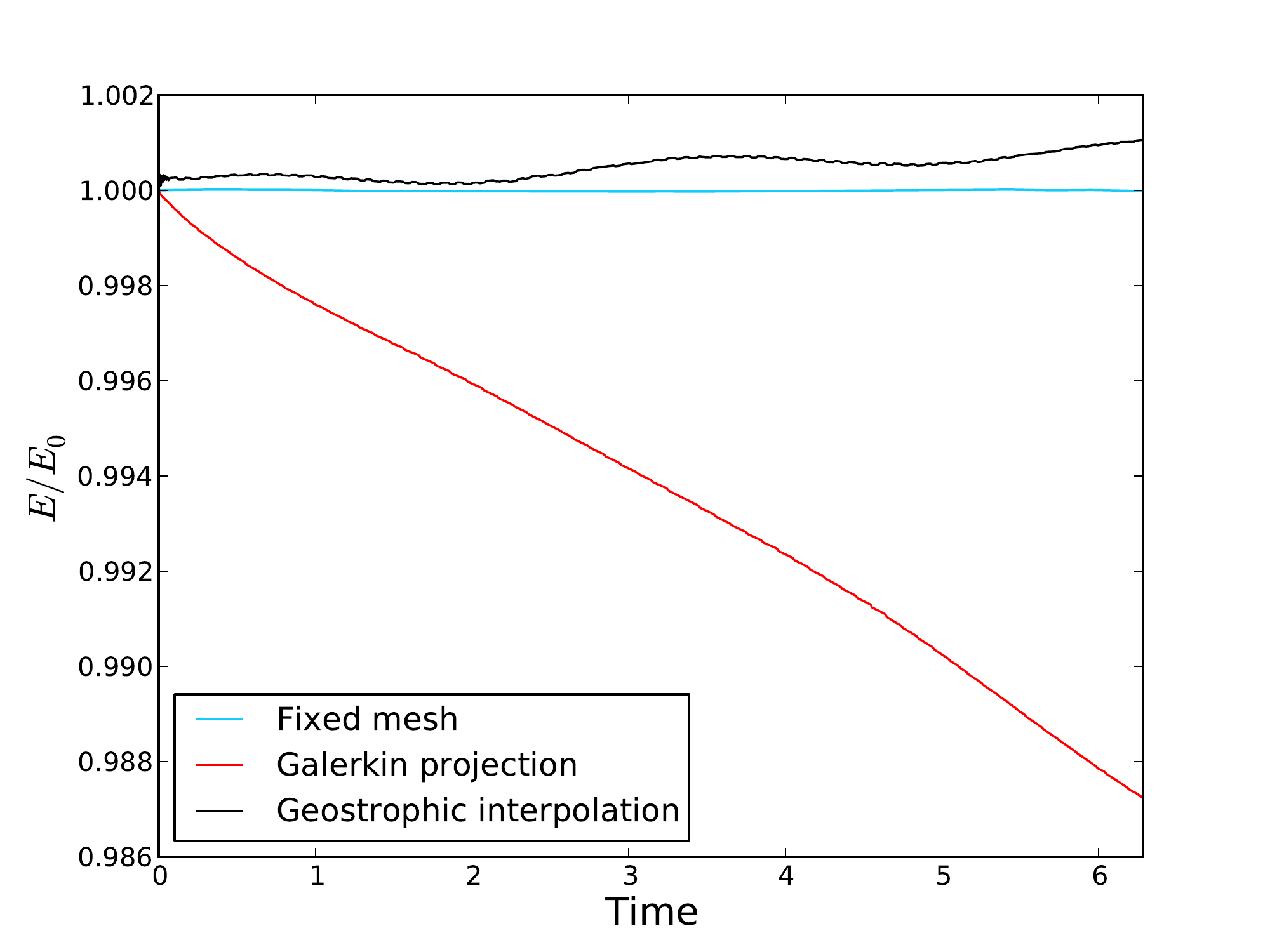}}
      \end{tabular}
    \end{tabular}
    \begin{centering}    
    \caption{The energy of the Kelvin wave test when interpolating between
             meshes using Galerkin projection and the geostrophic balance
             preserving interpolant, and when using a single fixed computational
             mesh.
             A: Kinetic energy.
             B: Potential energy.
             C: Total energy.}\label{fig:kw_energy}
  \end{centering}
\end{figure}

A discretisation of the linearised shallow-water equations conserves
energy if the layer thickness gradient matrix is, after
multiplication by some diagonal matrix, equal to the transpose of the
velocity divergence matrix \citep{ham2007}, and if the
implicit mid-point rule is used for timestepping \citep{leimkuhler2004}.
Hence the \chp\ spatial discretisation of the linearised shallow-water equations as presented
here conserves the total energy. The kinetic, potential, and total energy
of the system when using Galerkin projection, the geostrophic balance
preserving interpolant, and when using a single fixed computational mesh,
are shown in figure \ref{fig:kw_energy}. The fixed mesh simulation is observed to conserve the
total energy to within one part in $10^4$, with the relatively high error attributed to the
tolerances used for the linear solvers. The use of direct solvers, combined with more precision
robust calculation of the energy diagnostics, is expected to decrease this error. When interpolating between
meshes using Galerkin projection a systematic dissipation of both kinetic and potential energy
is observed, leading to a decrease in the total system energy of $1.3\%$ 
after $1000$ interpolations, at the end of the simulation. When using the
geostrophic balance preserving interpolant
a slight increase the potential energy is observed, leading
to an increase in the total system energy of $0.11\%$ after $1000$ interpolations.
While the geostrophic balance preserving interpolant is
not energy conserving, the change in system energy is, for this test,
more than an order of magnitude smaller than that observed when applying Galerkin
projection.

In further testing it was found that highly anisotropic elements intersecting the
domain boundary led to very poor results when using the geostrophic balance
preserving interpolant. This is likely due to significant interpolation errors
in the projection of $\Phi$ in this region, possibly as a result of the Dirichlet
boundary condition for $\Phi_c$ in the mesh-to-mesh Galerkin projection,
which pollutes the interpolated Coriolis acceleration. This problem was solved
by imposing constraints on the maximum element size for elements directly on the
boundary.

\subsection{Accuracy}\label{sect:accuracy}

Since the geostrophic balance preserving interpolant is composed of a
Galerkin discretisation of the Helmholtz decomposition of the
Coriolis acceleration followed by a donor-to-target
Galerkin projection of the decomposition, when using the \chp\ element pair
the interpolant is expected to be second order accurate for velocity and
third order accurate for layer thickness.

\begin{table}
  \begin{center}
  \begin{tabular}{| c | c |}
    \hline
    Donor mesh resolution ($x \times y$)& Target mesh resolution($x \times y$) \\
    \hline
    $ 24 \times  26$ & $ 26 \times  24$ \\
    $ 32 \times  35$ & $ 35 \times  32$ \\
    $ 48 \times  52$ & $ 52 \times  48$ \\
    $ 64 \times  70$ & $ 70 \times  64$ \\
    $ 96 \times 105$ & $105 \times  96$ \\
    $128 \times 130$ & $130 \times 128$ \\
    $192 \times 215$ & $215 \times 196$ \\
    \hline
  \end{tabular}\caption{Mesh resolutions used for the geostrophic balance
                        preserving interpolant convergence test. A mesh resolution
                        of $N \times M$ denotes a division in the $x$-direction
                        into $N$ sections, a division in the $y$-direction into $M$
                        sections, and the division of each resulting quadrilateral into
                        two triangles.}\label{tab:convergence_meshes}
  \end{center}
\end{table}

A series of structured triangular mesh pairs for a 2D unit square
$-0.5 \leq x \leq 0.5$ and $-0.5 \leq y \leq 0.5$ were generated
with resolutions in the $x$- and $y$-directions as given in table
\ref{tab:convergence_meshes}. A layer thickness of the form:

\begin{equation}
  \eta = \sin{\left( 2.5 \pi x \right)} \sin{\left( 2.5 \pi y \right)},
\end{equation}\label{eqn:convergence_layer_thickness}

and a velocity of the form:

\begin{equation}
  \vec{u} = -\vec{\hat{z}} \times \nabla \eta + \sin{\left( 0.5 x \right)} \vec{\hat{x}} + \sin{\left( 0.5 x \right)} \vec{\hat{y}},
\end{equation}\label{eqn:convergence_velocity}

were interpolated between the meshes in each pair using the geostrophic balance
preserving interpolant. The first term in \eqref{eqn:convergence_velocity}
corresponds to a flow that is, for $f = -1$, $g = 1$ and $H = 1$, in geostrophic
balance with the layer thickness \eqref{eqn:convergence_layer_thickness}. The
remaining terms correspond to a flow that cannot be balanced by any layer thickness.
In order to test for additional error introduced by the scalar potential $\Phi$
and layer thickness decomposition,
as per equation \eqref{eqn:p_decomp}, tests were conducted for a doubly
periodic and for a bounded domain. The \Ltwo\ errors
$\norm{\tilde{\eta}_B - \tilde{\eta}_A}{L^2}$,
$\norm{\tilde{u}_{x,B} - \tilde{u}_{x,A}}{L^2}$ and
$\norm{\tilde{u}_{y,B} - \tilde{u}_{y,A}}{L^2}$, were computed
explicitly via supermesh construction, as described in \citet{farrell2009a}.
For comparison the fields were also projected using Galerkin projection, giving
a measure of the quality of the geostrophic balance preserving interpolant
relative to the projection that is optimal in the \Ltwo\ norm.

\begin{figure}
  \begin{centering}
    \begin{tabular}{c c}
      \includegraphics[width=0.47\textwidth]{\addpwd{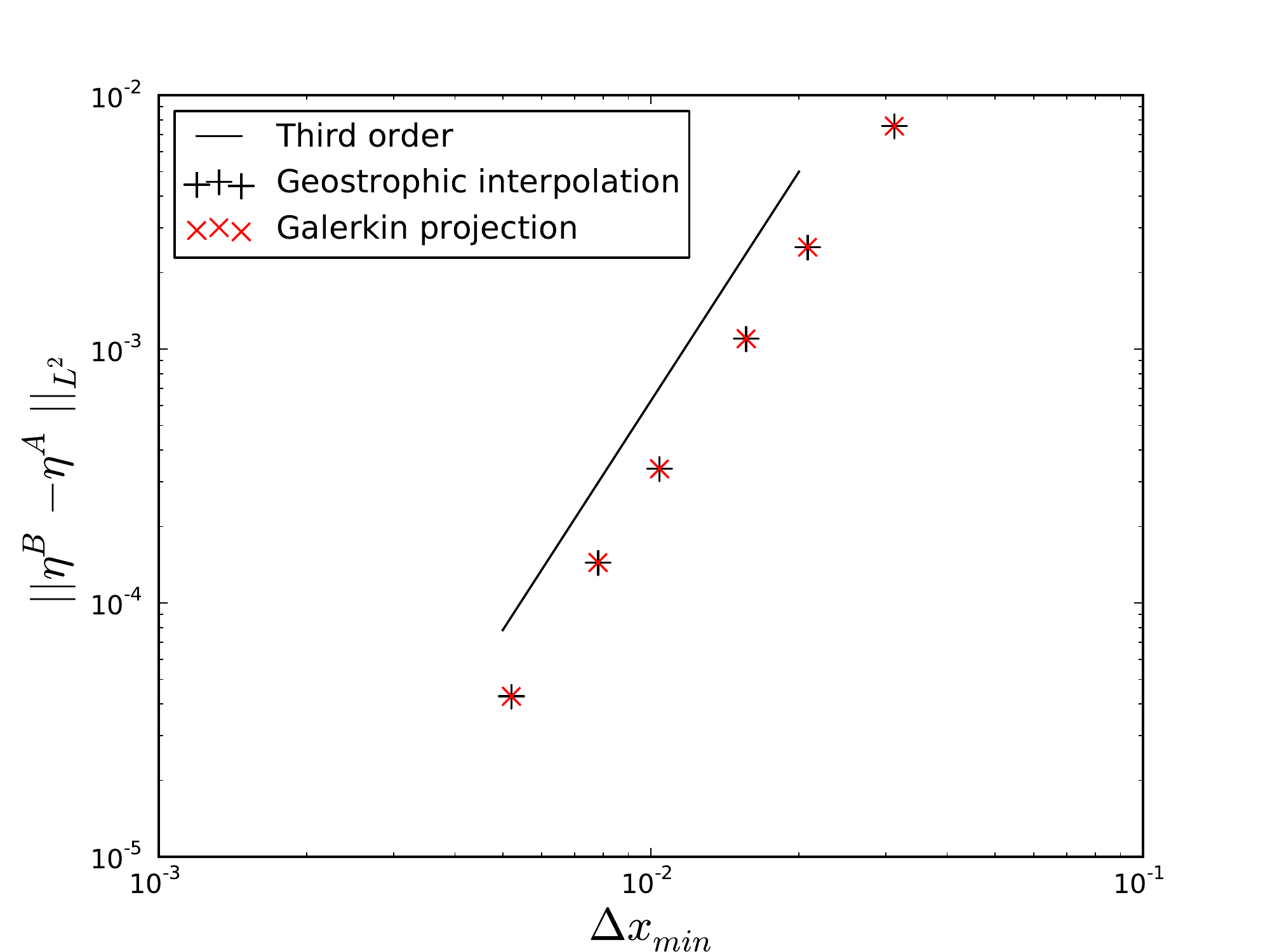}} &
      \includegraphics[width=0.47\textwidth]{\addpwd{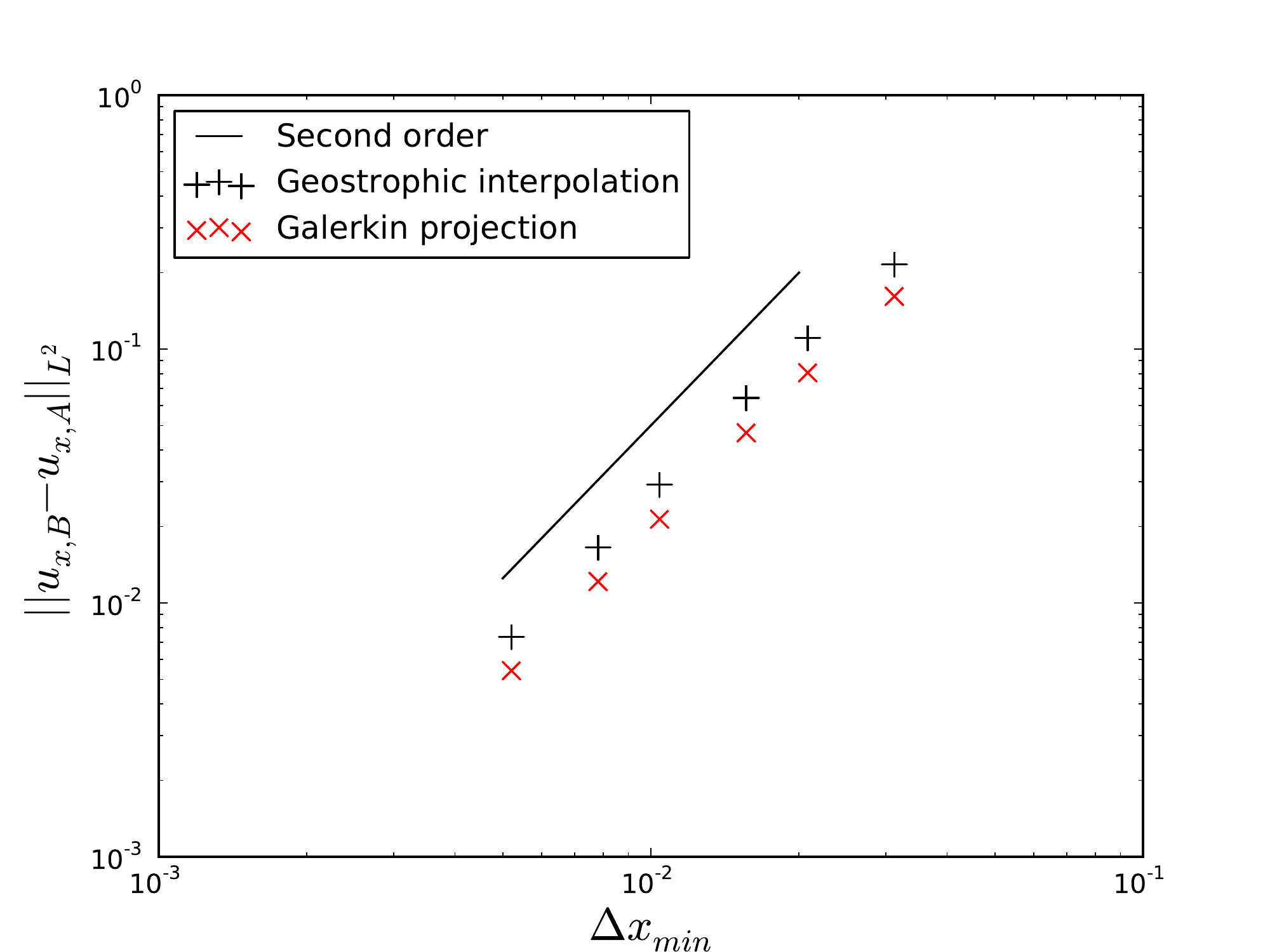}}
    \end{tabular}
    \caption{Convergence test for the geostrophic balance preserving interpolant
             in a doubly periodic domain, with Galerkin projection for
             comparison.
             Left: \Ltwo\ error in the layer thickness.
             Right: \Ltwo\ error in the $x$-component of velocity. The error in
             the $y$-component of velocity is similar.}\label{fig:convergence_dp}
  \end{centering}
\end{figure}

\begin{figure}
  \begin{centering}
    \begin{tabular}{c c}
      \includegraphics[width=0.47\textwidth]{\addpwd{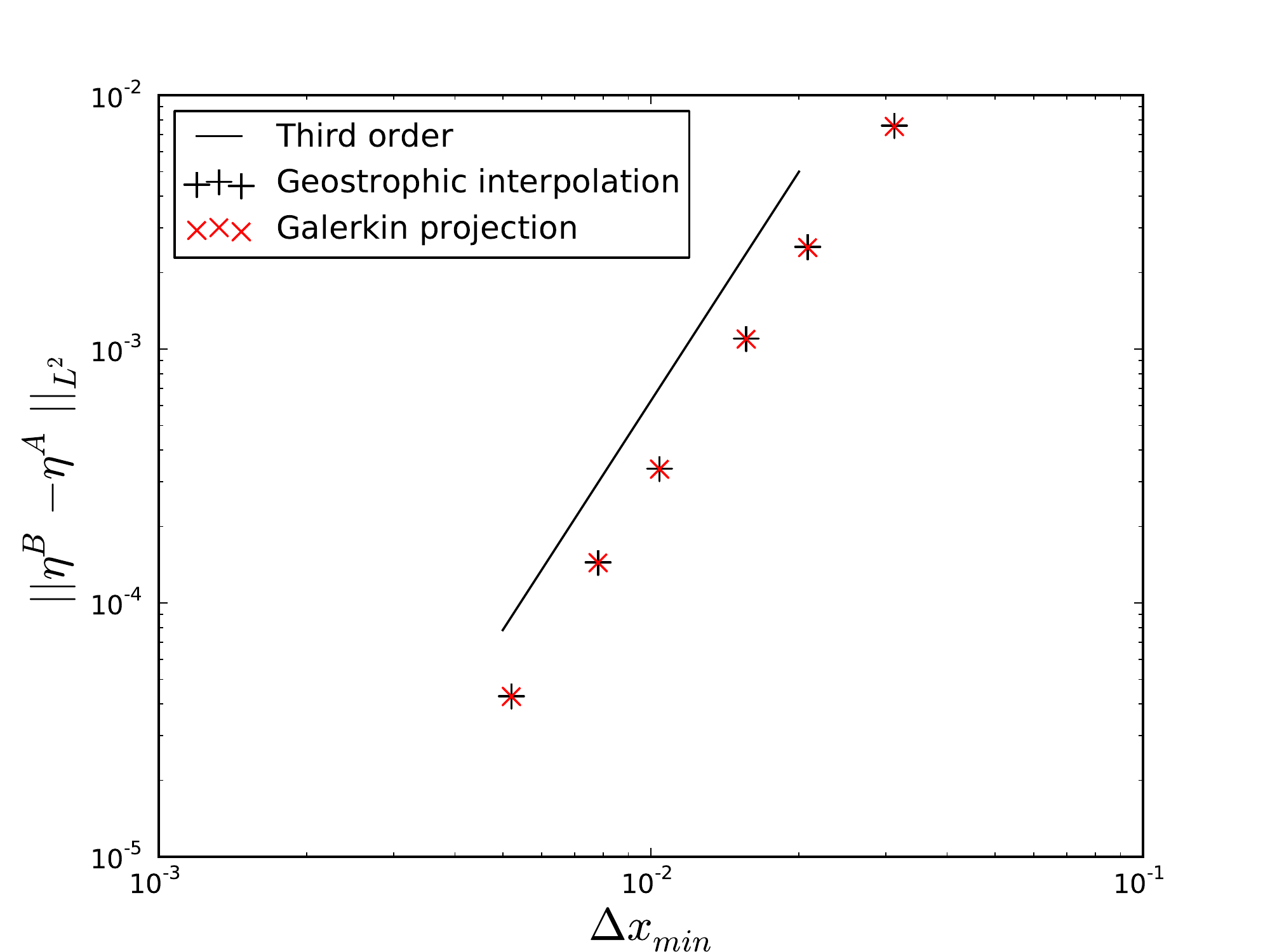}} &
      \includegraphics[width=0.47\textwidth]{\addpwd{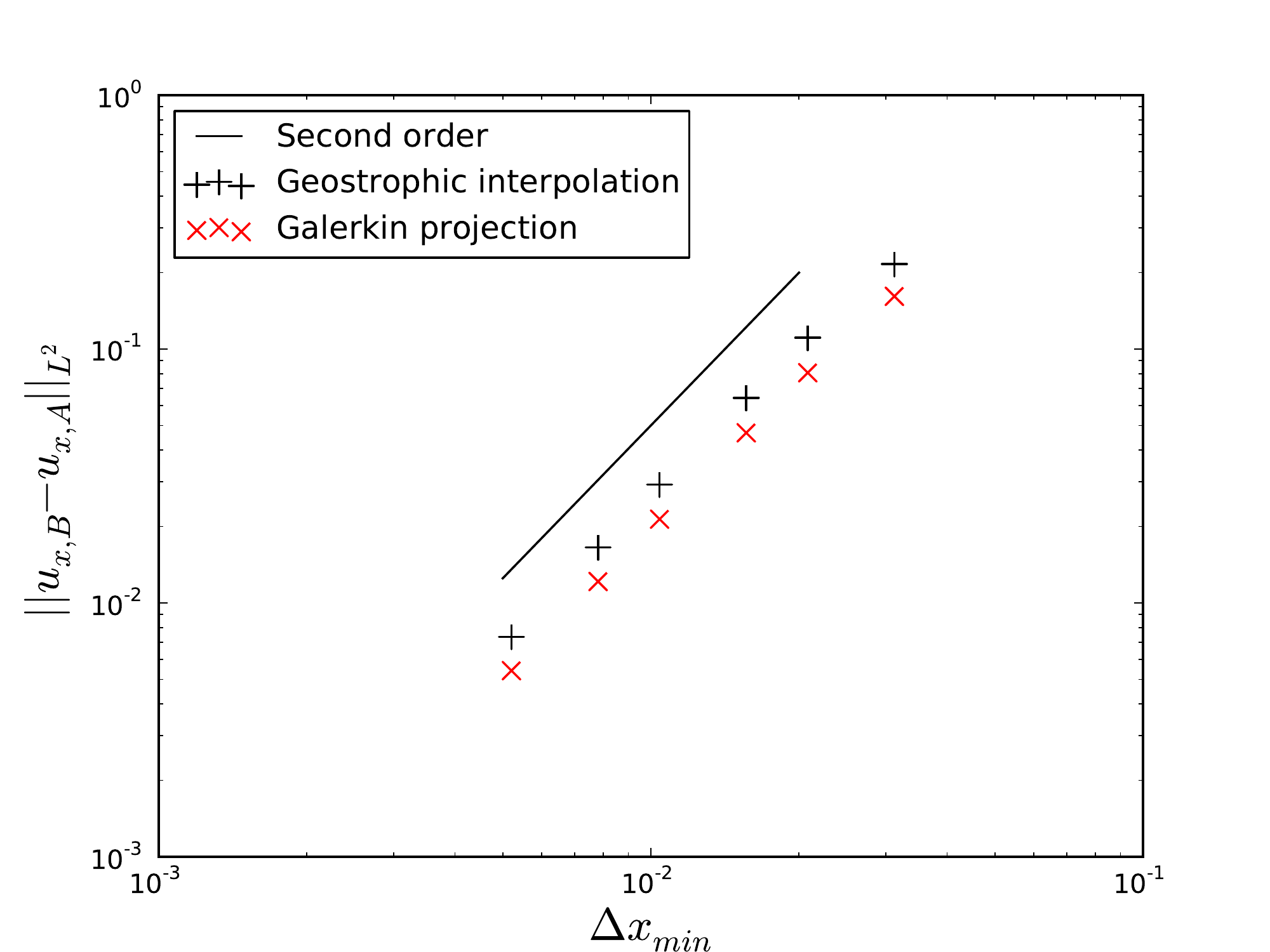}}
    \end{tabular}
    \caption{Convergence test for the geostrophic balance preserving interpolant
             in a bounded domain, with Galerkin projection for comparison.
             Left: \Ltwo\ error in the layer thickness.
             Right: \Ltwo\ error in the $x$-component of velocity. The error in
             the $y$-component of velocity is similar.}\label{fig:convergence_bounded}
  \end{centering}
\end{figure}

The resulting errors are shown for the doubly periodic domain in figure
\ref{fig:convergence_dp} and for the bounded domain in figure
\ref{fig:convergence_bounded}. The geostrophic balance
preserving interpolant is observed to be second order accurate for velocity
and third order accurate for layer thickness, as expected. For the
doubly periodic domain the average \Ltwo\ norm error for the geostrophic
balance preserving interpolant is observed to be $1.37$ times the optimal
value for velocity, and (since no layer thickness decomposition is applied in this case)
optimal for layer thickness. For the bounded domain
the error in velocity is not significantly changed, and the error in layer thickness is
increased to $1.005$ times the optimal value, indicating that the decomposition
of the scalar potential $\Phi_A$ and layer thickness $\tilde{\eta}_A$ 
introduces only a small additional error. For comparison, in \citet{farrell2009a}
collocation is found to give, for a field $\sin{x} + \cos{x}$, an
\Ltwo\ error that is $\sim2-2.5$ times the optimal value for piecewise linear
elements, and $\sim1.1$ times the optimal value for piecewise quadratic elements.

\section{Conclusions}

We have presented an interpolation method that, when applied to the
\chp\ discretisation of the linearised shallow-water equations on an
$f$-plane, guarantees that steady and geostrophically balanced states on the donor
mesh remain steady and geostrophically balanced after interpolation onto
an arbitrary target mesh. We have stress tested this balance preserving
property with highly anisotropic meshes and randomly initialised balanced
states (constrained to satisfy appropriate boundary conditions). We have
further demonstrated the utility of this interpolant for nearly balanced dynamics,
and quantified its accuracy in the \Ltwo\ norm.

A shortcoming of this approach, at least in the form presented, is
that is does not conserve energy. The Helmholtz decomposed
interpolation of Coriolis acceleration does not conserve kinetic energy or
potential energy. Despite this, the
change in energy when using the geostrophic balance preserving interpolant
was found to be more than an order of magnitude smaller than the energy dissipation
when using Galerkin projection. 
In addition to this, the interpolant does not locally conserve potential vorticity.
Geophysical flows are only in geostrophic balance to leading order, and a lack
of potential vorticity conservation could, when non-linear advection
is included in the fundamental equations, lead to higher order balance loss.
Potential vorticity decompositions could be considered where such a conservation
is desired \citep{staquet1989, holopainen1991, mcintyre2000}, although the benefit of such an approach,
bearing in mind that discretisations of the non-linear shallow-water equations are
not generally potential vorticity conserving, may be somewhat limited compared
to the benefit of leading order balance preservation.

The method has a natural extension to Navier-Stokes. For incompressible
Navier-Stokes any forcing that can be represented as the gradient of
a scalar field must be filtered by the pressure gradient, and hence the
interpolation of the Helmholtz decomposition of the Coriolis acceleration is
a balance preserving interpolant. Future work will concentrate on the
implementation of geostrophic balance preserving interpolation as part of
the Imperial College Ocean Model - an unstructured dynamic mesh adaptive ocean model. In particular,
we will investigate accurate preservation of geostrophic balance
when using velocity-pressure element pairs that do not satisfy optimal balance
properties, and the integration of methods used for accurate
balance representation for such element pairs \citep{ford2004, ford2004a, piggott2006, piggott2008a, fang2009}
into a balance preserving interpolant.
We will test how these can be used to propagate accurate balance representation
through arbitrary mesh adapts for meshes that are fully unstructured in all
three dimensions.
  
\section{Acknowledgements}

\noindent The authors wish to acknowledge support from the UK
Natural Environment Research Council (grants NE/C52101X/1,
NE/C51829X/1 and NE/H527032/1). The authors \linebreak would also like to thank
Prof. David P. Marshall and Dr. Hilary Weller for their comments and
suggestions in preparing this article.

\bibliographystyle{elsart-harv}
\bibliography{bibliography}

\end{document}